\documentclass{article}
\usepackage[utf8]{inputenc} 
\usepackage[T1]{fontenc}    
\usepackage{hyperref,url,booktabs}
\usepackage{amsfonts,amsmath,amssymb }      
\usepackage{nicefrac,microtype,graphicx,xcolor}      

\usepackage{cite} 

\usepackage{mathtools}

\usepackage{caption,subcaption}
\usepackage{fancyhdr,fancybox}
\usepackage[verbose=true,letterpaper,top=1in,bottom=1in,left=0.75in,right=0.75in]{geometry}
\usepackage{authblk}
\usepackage{arxiv}

\usepackage[export]{adjustbox}

\newcommand{\review}[1]{{\color{black} #1}}
\newcommand{\changed}[1]{{\review{#1}}}
\newcommand{\old}[1]{}

\title{Connecting the Dots: Range Expansions across Landscapes with Quenched Noise}
\author[a]{Jimmy Gonzalez Nu\~nez}
\author[b]{Jayson Paulose}
\author[c,d]{Wolfram M\"obius}
\makeatletter
\author[a]{Daniel A.~Beller}

\affil[a]{Department of Physics and Astronomy, Johns Hopkins University, Baltimore, MD}
\affil[b]{Department of Physics and Institute for Fundamental Science, University of Oregon, Eugene, OR}
\affil[c]{Living Systems Institute, Faculty of Health and Life Sciences, University of Exeter, Exeter, UK}
\affil[d]{Physics and Astronomy, Faculty of Environment, Science and Economy, University of Exeter, Exeter, UK}

\begin{document}
\maketitle

\begin{abstract}
    When biological populations expand into new territory, the  evolutionary outcomes can be strongly influenced by genetic drift, the random fluctuations in allele frequencies. Meanwhile, spatial variability in the environment can also significantly influence the competition between sub-populations vying for space. Little is known about the interplay of these intrinsic and extrinsic sources of noise in population dynamics: When does environmental heterogeneity dominate over genetic drift or vice versa, and what distinguishes their population genetics signatures? Here, in the context of neutral evolution, we examine the interplay between a population's intrinsic, demographic noise and an extrinsic, quenched random noise provided by a heterogeneous environment. Using a multi-species Eden model, we simulate a population expanding over a landscape with random variations in local growth rates and measure how this variability affects genealogical tree structure, and thus genetic diversity. We find that, for strong heterogeneity, the genetic makeup of the expansion front is to a great extent predetermined by the set of fastest paths through the environment. The landscape-dependent statistics of these optimal paths then supersede those of the population's intrinsic noise as the main determinant of evolutionary dynamics. Remarkably, the statistics for coalescence of genealogical lineages, derived from those deterministic paths, strongly resemble the statistics emerging from demographic noise alone in uniform landscapes. This cautions interpretations of coalescence statistics and raises new challenges for inferring past population dynamics.
\end{abstract}


\twocolumn
\newpage

\section*{Introduction}\label{sec:introduction}
Populations across many scales show evolutionary footprints of past range shifts. To properly interpret these footprints, it is important to understand how they originated. Range shifts are now understood to be common to many population histories \cite{Ricklefs2002}, from small\review{-}scale bacterial colonies \cite{Nadell2016,Xavier2007} and cancerous tissues \cite{Merino2016,Gidoin2018,Korolev2014} to large\review{-}scale species invasion in foreign biomes \cite{Fraser2015,Petren1996} and human migration \cite{Templeton2002,Henn2012,Sousa_2014,Peischl2016}; they thus play an important role in connecting evolution and ecology \cite{Excoffier2008,Ferriere2013}.

Large, well-mixed populations are insensitive to intrinsic, random genetic fluctuations, \text{i.e}., genetic drift.
In contrast, during range expansions, ``luck'' plays a significant role in providing rare mutations an opportunity to rise to high frequency and appreciably contribute to the expanding population \cite{Borer2020,Hallatschek2009,Fusco2016,Yu2021}. This increased influence of genetic drift arises because evolutionary competition becomes restricted 
to small effective population sizes at the boundaries between isogenic groups. As a result, local genetic diversity decreases rapidly with increasing expansion distance \cite{Hallatschek_2007,Excoffier_2009,Gralka2016,Lavrentovich2015}.  

In the absence of long-range dispersal,
genetic diversity at a traveling population front can be characterized by the motion of boundaries separating regions of similar genetic makeup, as well as by  branching genealogical lineage trees caused by repeated founder effects \cite{Neher2012}. When viewed backward-in-time, genetic dynamics are characterized by the intersections and coalescences of the population's lineages, traced from the population front back to the location of the initial populations. Along the way, pairs of related lineages sampled from the front eventually coalesce in a most recent common ancestor. The time since this coalescence event is then proportional to the number of accumulated (neutral) mutations expected to distinguish the genomes of the two sampled organisms, providing an important measure of genetic diversity \cite{Wakeley2008,Wilkins2002} and coupling the non-equilibrium statistics of propagating fronts to population genetics.

Little is known about how environmental heterogeneity interacts with this decay of genetic diversity and shapes the evolutionary outcome. Such an interplay has potential ecological and evolutionary consequences, e.g., the emergence of antibiotic resistance \cite{Baym_2016} and altered species invasion dynamics \cite{With1997,With2002}.
Some progress has been made in understanding statistics of relatedness across the landscape in bounded environments \cite{Nullmeier_2013}, for single obstacles to the expansion \cite{Moebius2015}, and for environments with curved surfaces \cite{Beller2018}. 
In particular, individual obstacles, which suppress growth locally, act similarly to bumps in surface topography by focusing the expansion front inward to a cusp \cite{Moebius2015,Beller2018}. Landscapes of randomly placed obstacles create pinning sites for the population front and increase the importance of chance relative to selective fitness in determining the genetic makeup of the expansion \cite{Gralka2019}. 

However, it remains poorly understood how a heterogeneous growth landscape influences the genetic structure of expanding populations. To study this interplay, we examine the effects of environmental heterogeneity on a minimal model of population growth, a multi-species variant of the Eden model \cite{eden1961two}, in which the front of the range expansion is a propagating, roughening interface in the Kardar-Parisi-Zhang (KPZ) universality class \cite{Kardar1986}. Motivated by experiments on bacterial and yeast colonies grown on agar plates with limited nutrients \cite{Hallatschek_2007,Hallatschek2009,Hallatschek2010,Gralka2016, Korolev_2011}, the evolutionary dynamics of this model in uniform environments has been extensively studied, including phenomena such as fitness collapse \cite{Castillo2020,Lavrentovich2015a}, fixation \cite{Krishnan2019}, gene surfing \cite{Hallatschek2008,Gralka2016}, and lineage coalescence \cite{Chu_2019,Gralka2019}. 

In this work, we couple the Eden model's range expansion to a landscape of \emph{hotspots} where the population expands faster locally. The effects of  hotspots, both individually and in randomly placed collections, on the population dynamics have recently been characterized by employing an analogy of light rays passing through a medium of regions with locally decreased index of refraction \cite{Moebius2021}. We demonstrate that for strong heterogeneity, despite appearing stochastic, the population's lineage structure is dominated by geometrically determined paths of least travel time, and  these paths predict which initial sub-populations will dominate the late stages of the expansion. Strikingly, we find that strong environmental noise can produce statistics in the population's genetic composition that mimic scaling due to demographic noise on uniform landscapes, despite making the genetic outcomes nearly deterministic.

\section*{Model}\label{sec:methodology}
\begin{figure}[tb]
    \centering
    \includegraphics[width=0.49\textwidth]{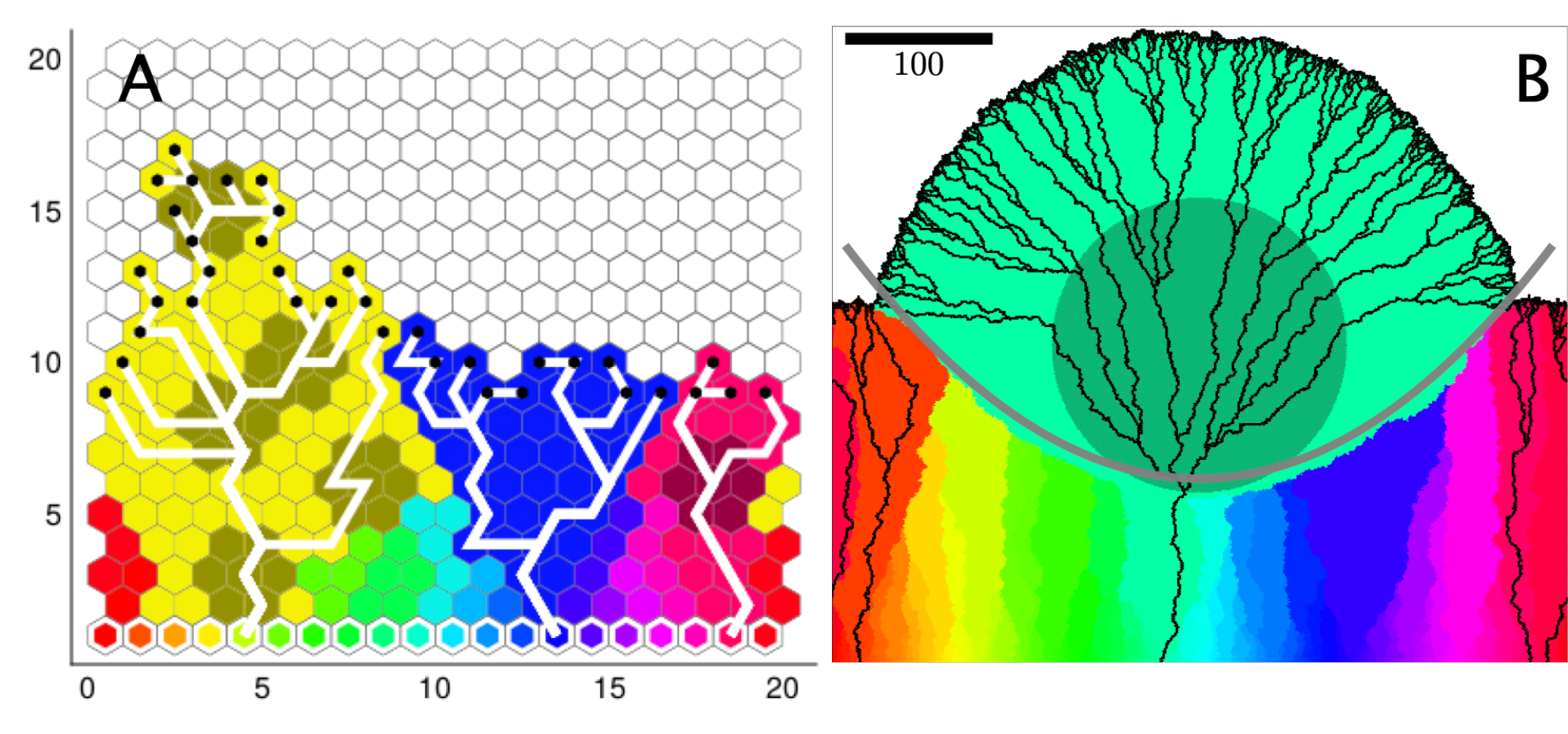}
    \caption{Illustration of Eden model for a range expansion on a heterogeneous landscape of hotspots with a linear initial condition (white-outlined hexagon markers), and with each color representing a distinct ancestral deme.  Gray sites make up the \emph{hotspots}, regions of increased reproduction rate. (A) Simulation \review{snapshot} showing lineages (white lines) at the scale of the lattice and for intensity $I=8$. Black dots indicate the population front. (B) Simulation \review{snapshot} showing the effect of a single hotspot on the lineages (black lines) and population front, using intensity $I=10$. Parabolic approximation to the sector boundaries induced by the hotspot is shown in  gray \review{(Eq.~\ref{eq:parabola}; see Eq.~1 of \textbf{SI appendix} for full expression)}.}
    \label{fig:latticeHotspots}
\end{figure}

Our simulated population is arranged on a hexagonal grid, with each filled grid site containing a \emph{deme} (well-mixed sub-population) that can grow or expand into neighboring sites (Fig.~\ref{fig:latticeHotspots}\textit{A}). In the context of a continuous population, this corresponds to the regime where growth to local carrying capacity occurs faster than migration. For this reason,  the genetic character of each deme is assumed to be uniquely determined by the first individual to arrive. We consider reproduction of demes to occur only at the population boundary, \textit{i.e.}, no replacement of an occupied deme is possible, leading to a population composition that is ``frozen'' in time. 
In this scenario, we can use a single identifier to characterize the local genetic composition, which we represent with a distinct color (Fig.~\ref{fig:latticeHotspots}\textit{A}). 

Our initial condition consists of a single, filled edge of $L$ sites on our hexagonal grid (Fig.~\ref{fig:latticeHotspots}\textit{A}). We choose each initially occupied site to be identified with one of $L$ colors with all demes sharing equal intrinsic fitness (neutral evolution). We employ asynchronous reproduction rules known to reproduce the scaling (discussed in more detail later) of domain boundary fluctuations \cite{Hallatschek_2007} and lineage lateral motion \cite{Gralka2016} often seen in microbial experiments, as well as many classes of systems with kinetically roughed fronts \cite{Halpin_Healy_1995}.
 
These reproduction rules consist of (1) identifying the population front by all demes near an empty grid site, (2) randomly selecting one deme at the population front according to an implementation of the Gillespie algorithm \cite{Gillespie1976,Cai2009}, and  (3) randomly selecting one neighboring grid site with uniform probability to establish a new deme of the same color as the replicating deme. For neutral evolution in uniform environments, this amounts to selecting individual demes at the front with equal probability and copying their color to a random adjacent empty site, \textit{c.f.}\ Eden model \cite{eden1961two}, type C \cite{Jullien1985}. Because color is inherited at replication, this process leads to single-colored regions, each occupied by descendants of one individual from the original population (Fig.~\ref{fig:latticeHotspots}\textit{A}). By tracking replications, we can also visuali\review{z}e the ancestral relationships between sites. Here, we focus on the ancestral history of individuals at the front, which are represented by lineages that coalesce when viewed backwards in time (white lines in Fig.~\ref{fig:latticeHotspots}\textit{A}).

Environmental heterogeneity in our system takes the form of a fixed distribution of disk-shaped ``hotspots'', which are regions of increased local reproduction rate (Fig.~\ref{fig:latticeHotspots}). Thus, for heterogeneous environments, step (2) is modified first by assigning to each grid site $(i,j)$ a reproduction rate, 
\begin{equation}\label{eq:growth_rates}
    r(i,j) = \begin{cases} 
        r_h & \text{if $(i,j)$ in a hotspot,}\\
        r_0 & \text{otherwise,}
    \end{cases}
\end{equation}
where $r_h, r_0$ are the reproduction rates inside and outside of a hotspot, respectively. We implement a version of the Gillespie algorithm \cite{Gillespie1976} to update the simulation time and select the next deme to reproduce\review{;} see \textbf{Materials and Methods}.
A single disk-shaped hotspot leads to an expanding population bulge \review{(Fig.~\ref{fig:latticeHotspots}\textit{B})}, as characterized in Ref.~\cite{Moebius2021}.
Our structured environment takes the form of a ``landscape'' of hotspots with Poisson-distributed centers. The hotspots include all lattice sites centered within a distance $R$ of any hotspot center, and with overlap allowed between neighboring hotspots. This gives rise to three system parameters: the hotspot radius $R$, the area fraction $\phi$ of hotspots, and the ratio of replication rates $r_h/ r_0$. The latter parameter can be recast as the hotspot intensity, $I=({r_h}/{r_0}) - 1$. Because the replication rate is proportional to front speed, hotspot intensity can be rewritten as $I=({v_h}/{v_0}) - 1$. \review{Here,} $v_h$ and $v_0$ \review{are the} front speed within and outside the hotspots, respectively\review{, when measured on spatial scales sufficiently large that an effective front speed is meaningful}.

\begin{figure}[htb]
     \centering
     \includegraphics[width=0.49\textwidth]{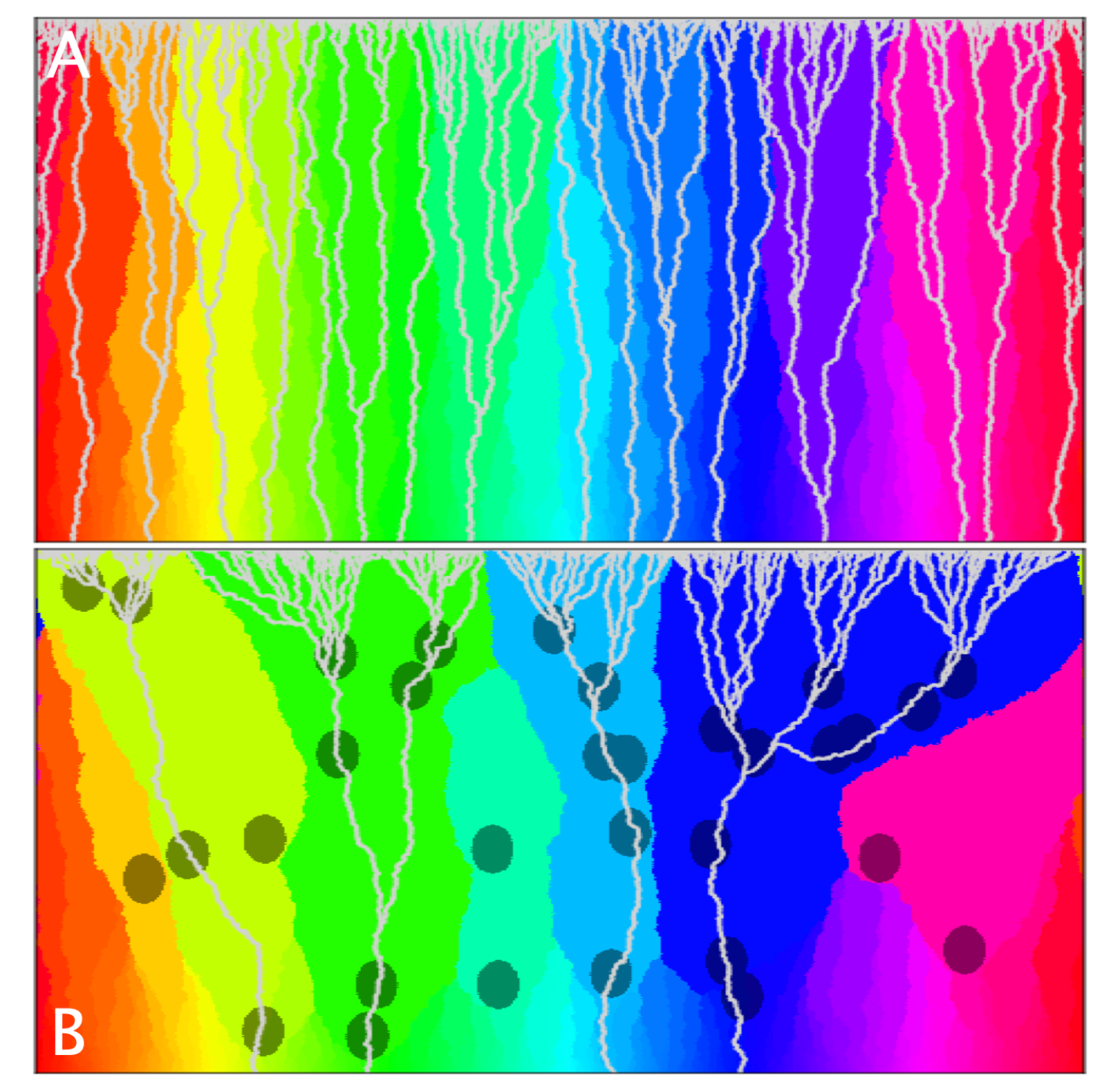}
    \caption{Illustration of a range expansion from a linear initial population of 2000 demes wide grown to 1000 demes tall showing sector coarsening (colors) and lineage structure (light gray trails), (A) without hotspots, and (B) with hotspots (dark gray disks) at an area fraction $\phi=0.09$, radius $R=20$ and intensity $I=8$.}
    \label{fig:landscapeView}
\end{figure}

Simulations are performed on a hexagonal grid with periodic boundaries along the width direction, and beginning with initial populations of $L$ \review{$=2000$} demes. \review{For each value of hotspot intensity and hotspot area fraction, we conducted 100 simulation runs with distinct random seeds over each of 20 different landscapes. Populations are}
grown only to a height of $L/2$ demes to reduce finite size effects in genetic measures by preventing lineages from self-interacting. Specifically, viewed forward in time, replicating demes which happen to establish new demes consistently in one direction will have descendent demes that have been displaced at most a  distance equal to half the system width. Viewed backward-in-time, then, lineages constructed by tracing these replication events backwards from demes at the population front toward a founding ancestor do not have enough time to wrap through the periodic boundaries onto themselves. 

\section*{Results}\label{sec:findings}
\subsection*{Effect of  a single-hotspot on genetic diversity}
Expansion of the population on the lattice leads to a loss of genetic diversity, which can be seen in
the decrease in the number of distinct colors with increasing height, Fig.~\ref{fig:latticeHotspots}\textit{A}.  Occasionally a deme of a particular color will be surrounded by demes of other colors,  preventing it from establishing new descendants; two sector boundaries then merge into one. Since a merger of boundaries \review{disconnects a genetic sector from the front and thus coincides with} the termination of \review{that sector's} lineages, sector coarsening is strongly coupled to lineage structure. That is, \review{a decrease} in the number of genetic sectors also correspond\review{s} to a decrease in the number of founder-population ancestors that ``survive'', \textit{i.e.} that remain present as distinct roots of lineage trees (white lines in Fig.~\ref{fig:latticeHotspots}\textit{A}, gray lines in Fig.~\ref{fig:landscapeView}) traced backward from the population front. This has been studied extensively in uniform environments\review{, where, for front-roughening dynamics in the KPZ universality class,} the number of surviving sectors \review{decreases with mean expansion distance $h$ as $n_s \sim h^{-2/3}$\cite{Saito1995}} and \review{lineages coalesce in reverse-time $\tau$ such that the number of distinct lineages scales as $\sim \tau^{-2/3}$}
 \cite{Gralka2016}.

Fig.~\ref{fig:latticeHotspots}\textit{B} shows a range expansion at a scale much larger than the lattice spacing with a single hotspot at the center of the landscape with hotspot intensity $I=10$. \review{The outward-biased motion of sector boundaries induced by the hotspot corresponds to descendants of demes at the population front that enter the hotspot being directed to the hotspot periphery. This also leads to a} rapid reduction in the number of sectors in the vicinity of the hotspot\review{, meaning} that those unlucky few lineages that pass near the hotspot without entering it will very likely be cut off by the sectors influenced by the hotspot. An animation of this effect is provided in Supplementary Movie S1. The resulting sector boundary in Fig.~\ref{fig:latticeHotspots}\textit{B} (gray curve) is well described by a parabola \review{whose coefficient for the quadratic term is given by } 
\begin{equation}\label{eq:parabola}
    a = (1 + I) / (4 IR).
\end{equation}
\review{The parabola with this coefficient emerges as the set of intersection points of the unperturbed population front and a circular population front originating at the hotspot center, which have been found to heuristically describe front propagation outside hotspots \cite{Moebius2021}}. The detailed form for the parabolic sector boundary including the position of the vertex and relation between front shape and sector boundary is given in \textbf{SI Appendix}.

For lineages, the hotspot acts like a diverging lens (viewed from bottom to top, Fig. \ref{fig:latticeHotspots}\textit{B}). This is in contrast to obstacles and topographic bumps where lineages are guided around the environmental feature \review{analogously to the focusing of light by} a converging lens \cite{Moebius2015,Beller2018}. For topographic bumps it was shown that average lineage trajectories are well predicted by analytically calculated geodesic paths \cite{Beller2018}, \textit{i.e.}, the paths of least time. In this work, we therefore expect fastest paths through a landscape of many hotspots to provide useful predictions for trajectories of surviving lineages. We investigate this hypothesis in detail below, in \textbf{Least Time Principle and Lineage Structure} and \textbf{Meandering of Lineages and of Fastest Paths}.

\subsection*{Many-hotspot effects on diversity} \label{subsec:many-hotspot}
While a single hotspot has a qualitatively simple geometrical effect on front geometry, sector coarsening, and shape of lineages, the situation is complicated in the presence of many hotspots. It is known that the shape of the advancing population front remains well described by geometric optics calculations for paths of least time \cite{Moebius2021}. Here, we investigate the consequences that landscapes of many hotspots have on the population's genetic diversity.

Example simulations for a range expansion through a uniform landscape ($\phi=0$), and a landscape with hotspot area fraction $\phi=0.09$, hotspot size $R=20$, and intensity $I=8$ are shown in Figures \ref{fig:landscapeView}\textit{A,B} respectively. We have provided an animation for a range expansion in a heterogeneous landscape in Supplementary Movie S2. Comparing these two figures shows that landscapes of many hotspots qualitatively hasten genetic sector coarsening, resulting in a corresponding drop in the number of surviving lineages, and leading to greater lateral meandering in both lineages and sector boundaries.

\subsubsection*{Sector coarsening\label{sec:sectorcoarsening}} 
\begin{figure}[tb]
    \centering
    \includegraphics[width=0.49\textwidth]{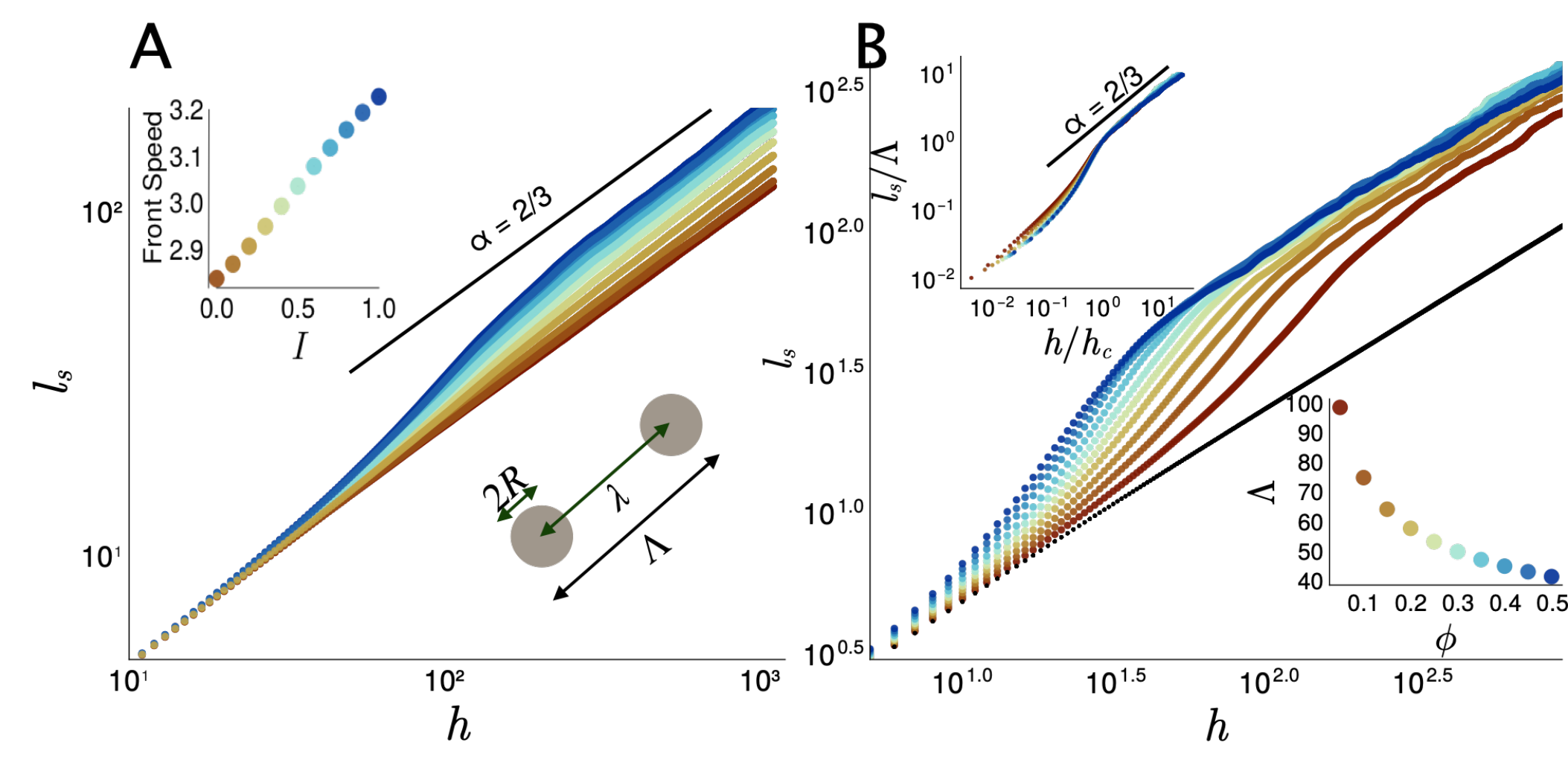}
    \caption{\review{Log-log plot of the (A) average genetic lateral sector size $l_s$ plotted against expansion distance $h$  at varying hotspot intensity $I$ and fixed area fraction $\phi=0.1$, (B) sector growth for varying hotspot area fraction $\phi$ at fixed intensity $I = 8$; black markers indicate a uniform landscape ($\phi=0$). Top-left insets show (A) population front mean speed dependence on hotspot intensity, (B) rescaled sector size vs rescaled height $h / h_c$ (defined in the main text) for $\phi>0$. The hotspot radius is fixed at $R=10$. Bottom-right insets show (A) schematic of hotspot typical outer-edge-to-outer-edge distance $\Lambda$, (B) dependence of this distance on hotspot area fraction, $\Lambda(\phi)$.}}
    \label{fig:sectorSizes}
\end{figure}

To study the decay in the number $n_s$ of genetic sectors, we measure how the average sector width $l_s \sim 1 / n_s$ grows as a function of the average front height (expansion distance) $h$ (Fig.~\ref{fig:sectorSizes}).
In the absence of hotspots, there is experimental \cite{Hallatschek_2007} and numerical \cite{Saito1995} 
evidence of a scaling relation $l_s \sim h^{\alpha}$, with $h$ being expansion distance from initial population and approximate KPZ scaling exponent of $\alpha \approx 2/3$. In agreement with this expectation, we find that \review{uniform landscapes} ($I=0$ in Figure \ref{fig:sectorSizes}\textit{A}, data with black markers ($\phi=0$) in Figure \ref{fig:sectorSizes}\textit{B}) exhibit KPZ scaling $l_s \sim h^{\alpha}$, $\alpha=2/3$ for all heights beyond some initial height on the order of 10 demes.

We explore how $l_s$ changes with two parameters describing the many-hotspot landscape: hotspot intensity $I$ (Fig.~\ref{fig:sectorSizes}\textit{A}) and hotspot area fraction $\phi$ (Fig.~\ref{fig:sectorSizes}\textit{B}). Tuning $I$ from 0 to 1 results in a  transient regime at heights $ h \gtrsim 100$ which deviates from the uniform\review{-}landscape case, and where sector size grows faster than $\sim h^{2/3}$. After this transient regime, at high $h$, sector growth returns to the KPZ scaling $\alpha=2/3$, but with a prefactor that is greater for larger hotspot intensities (Fig.~\ref{fig:sectorSizes}\textit{A}). We see similar behavior when varying $\phi$, with the return to the KPZ regime at high $h$ occurring slightly faster for higher area fractions (Fig.~\ref{fig:sectorSizes}\textit{B}).

For landscapes of many hotspots, previous work characterized front propagation by an effective front speed, approximating the heterogeneous landscape as a uniform, effective medium \cite{Moebius2021}. 
This effective front speed increases with both hotspot intensity $I$ (Fig.~\ref{fig:sectorSizes}\textit{A} top-left inset) and area fraction $\phi$ (\textbf{SI Appendix} Fig.~S4). In this light, \review{it is tempting to view} the return to KPZ scaling at large $h$ as signalling a regime where \review{sector dynamics are well-described by an effective medium picture} at expansion distances larger than some crossover length. \review{As we will see below, the large-scale dynamics is not truly that of an effective medium because the wandering of lineages in this regime is dominated by the geometry of the environment, not by demographic noise.
Nonetheless, we can better} understand the transient behavior and the return to KPZ scaling at large expansion distance \review{by defining} a length scale for the typical hotspot center-to-center distance, $\lambda \equiv 1/\sqrt{\rho}$, where $\rho$ is the hotspot number density, \review{i.e., the number of hotspots per unit area}. For a random distribution of disks permitted to overlap, this length is related to the hotspot area fraction $\phi$ and radius $R$ through \cite{Xia1988,Torquato2002}
\begin{align}\label{eq:lambda}
    \lambda(\phi,R) \equiv \frac{1}{\sqrt{\rho}} = \sqrt{\frac{\pi R^2}{-\ln(1 - \phi)}}.
\end{align}

Empirically, we find that the crossover length is well described by 
the typical outer-edge-to-outer edge distance $\Lambda \equiv \lambda + 2R$ of two nearby hotspots (illustrated in Fig.~\ref{fig:sectorSizes}\textit{A} bottom-right inset).
Values for $\Lambda$ calculated using Eq.~\ref{eq:lambda} are shown in Fig.~\ref{fig:sectorSizes}\textit{B} bottom-right inset for corresponding values of $\phi$. \review{Those $\Lambda$ values provide fairly good approximations for the crossover length. Indeed, rescaling the sector size by the hotspot separation length scale, $l_s / \Lambda$, and the expansion distance to $h/h_c$, with $h_c$ being the height at which the sector size reaches $\Lambda$, collapses the late-time sector size data, with the transition now occurring at $(1,1)$ on the rescaled axes for all $\phi$ (Fig.~\ref{fig:sectorSizes}\textit{B} top-left inset). This reveals that sector size growth returns to KPZ scaling at sufficiently large sector sizes, relative to the environmental length scale $\Lambda$.}

\subsubsection*{Environmentally Pinned Lineages}
\begin{figure}[b]
    \centering
    \includegraphics[width=0.5\textwidth]{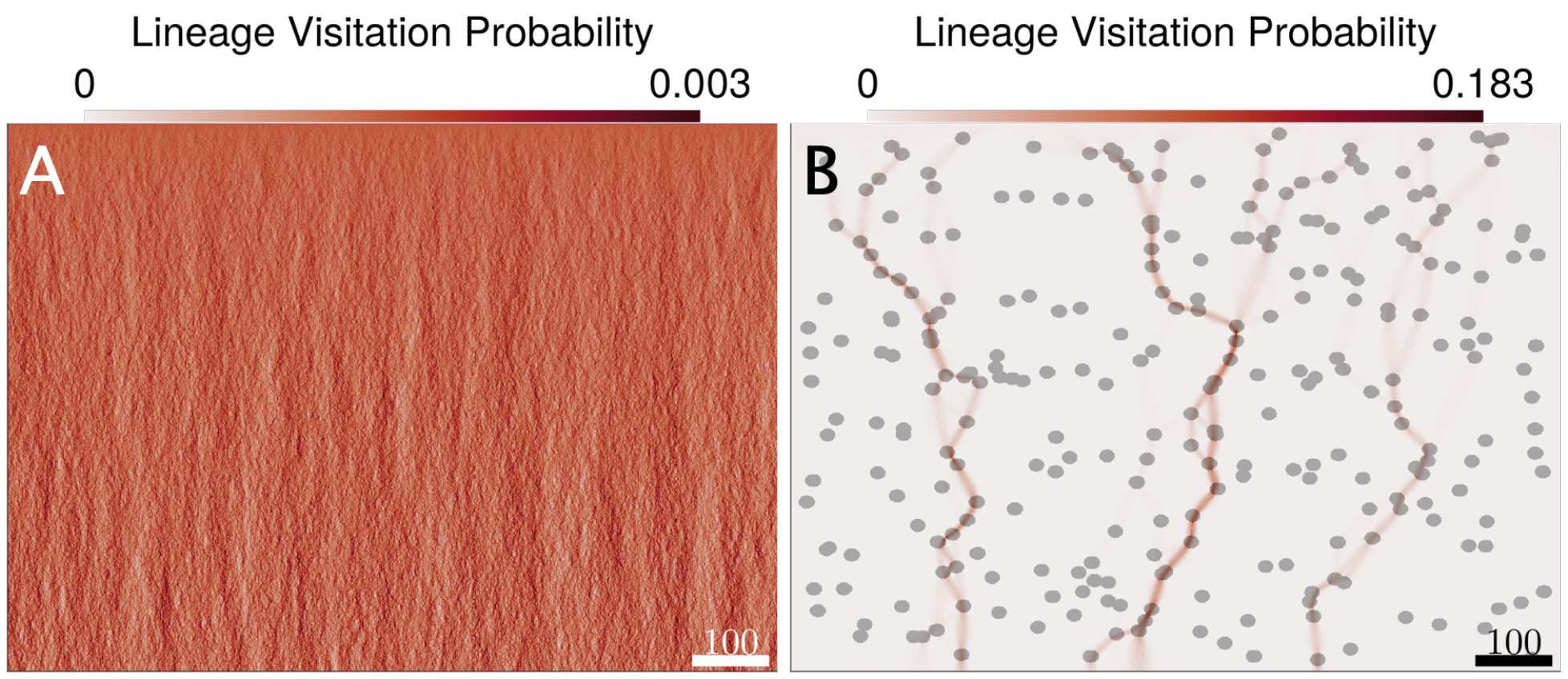}
    \caption{Lineage site-visitation probability map generated from an ensemble of 1200 simulations with (A) no hotspots and (B) a fixed landscape of hotspots with an intensity $I=3$, size $R = 10$, and area fraction $\phi = 0.09$.}
    \label{fig:hotspot_pinning}
\end{figure}

As described above, at sufficiently large expansion distances, genetic sector coarsening in many-hotspot landscapes scales similarly to that on uniform landscapes. Since the sector dynamics and lineage structure are related, it would be reasonable to expect similar stochastic behavior for lineages. Indeed, individual simulations, such as Fig.~\ref{fig:landscapeView}\textit{B}, may at first seem to indicate that growth through a landscape of hotspots produces lineage trees similar to those in uniform environments, although with sector coarsening accelerated by hotspots (see also Supplementary Movie S2). 

However, at close inspection, comparison of simulation runs from different initial random seeds but within the same landscape and initial condition reveals a qualitatively distinct feature of range expansions in landscapes with quenched random noise: When the hotspot intensity is sufficiently large, a large proportion of the lineages follow certain highly traversed paths through a given landscape, traveling through a small, consistent subset of the hotspots. We dub this phenomenon \emph{lineage pinning}. This effect is seen by averaging the lineage spatial positions over many simulations so that each lattice site records a probability of being visited by a surviving lineage. Fig.~\ref{fig:hotspot_pinning}\textit{A} shows such a lineage site-visitation probability map for a uniform environment, with darker colors indicating more commonly visited positions. The observed texture arises from a finite ensemble of 1200 simulations; for an infinite ensemble in a uniform environment, all positions at a given height would be visited equally, by symmetry. In a heterogeneous landscape, a very different pattern emerges in Fig.~\ref{fig:hotspot_pinning}\textit{B}\review{,} where the presence of hotspots induces a clear structure of frequently traversed \review{lineage} positions passing through a subset of the hotspots.

With the front at the end of the simulation consisting of $L$ demes by construction, the average lineage tree pattern in Fig.~\ref{fig:hotspot_pinning}\textit{B} indicates that there is a funneling of lineages to a reduced number of paths through the landscape, ultimately terminating on a finite subset of ancestral demes (located at height zero). This suggests that structured environments induce favorable paths for lineages and thus impart a deterministic component to the evolutionary dynamics: the pattern of hotspots biases the genetic composition of the expansion front in favor of certain \review{ancestral} demes. In the forward-time view, this also implies deterministic contributions to genetic sector survival and dynamics of sector boundaries. The onset of lineage pinning with increasing hotspot intensity $I$ is demonstrated in Supplementary Movie S3, which shows the lineage visitation probability as $I$ is varied while the hotspot landscape is otherwise held fixed. As $I$ is increased, paths of high visitation probability emerge and become narrower as they accumulate more of the total probability. 

\begin{figure}[tb]
    \centering
    \includegraphics[width=0.49\textwidth]{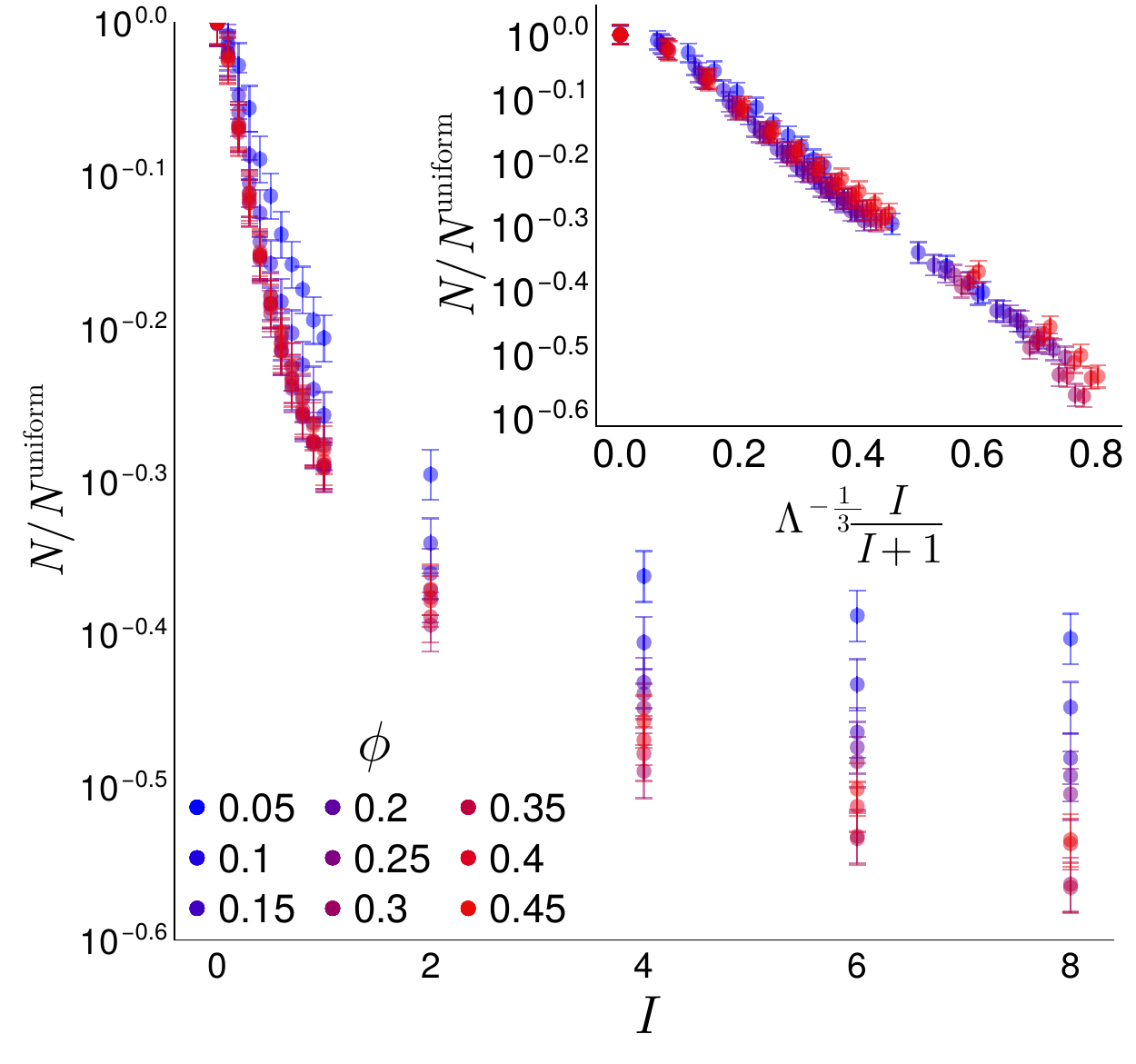}
    \caption{\review{Normalized number of surviving lineages after a range expansion as a function of hotspot intensity, $I$, for fixed radius $R=10$ with each data point generated from 20 different landscapes each with 100 simulation runs. Inset shows a possible scaling collapse with pinning parameter $\Lambda^{-1/3} I / (I+1)$.
       Error bars represent standard deviations.}}
    \label{fig:surviving_ancestors}
\end{figure}

\review{
A characteristic of lineage pinning is that, as hotspot intensity is increased on a given landscape, the lineage site-visitation probability distribution ``condenses'' gradually onto a few paths (Supplementary Movie S3). These most probable lineage paths tend to remain stable in position while increasing in probability as $I$ increases.  \review{This means that survival probability of the progeny of the ancestral demes becomes concentrated in a landscape-defined subset of ancestral demes as the hotspot intensity increases, as shown in Fig.~\ref{fig:heatmapLineages}\textit{D}.}

Lineage pinning accelerates the decay of genetic diversity, as seen in a reduction in the number of surviving lineages, \textit{i.e.}, ancestral demes contributing to the front's genetic makeup: 
As one summary statistic from these simulation ensembles, Fig.~\ref{fig:surviving_ancestors} displays the mean number of surviving lineages (number of \review{lineage} tree roots), normalized by the corresponding number in the absence of hotspots, as a function of hotspot intensity. Values are smaller than 1, indicating that lineage pinning by the environment reduces the number of distinct roots exhibited by a given lineage tree.

The $\phi$-dependence of this accelerated decay of surviving lineages can be approximately collapsed by plotting the normalized number of surviving lineages against $\Lambda^{-1/3} I / (I+1)$
with $\Lambda = \lambda + 2R$ and $\lambda$ related to $\phi$ and $R$ through Eq.~\ref{eq:lambda}. The $I$-dependence for this collapse is motivated by Eq.~\ref{eq:parabola} and the power of $\Lambda$
is found empirically, as shown in  Fig.~\ref{fig:surviving_ancestors} (Inset). This rescaling collapse holds only for densities less than $0.5$, presumably due to the hotspot separation length scale rapidly approaching the hotspot diameter near $\phi \approx 0.55$}.

\subsubsection*{Least Time Principle and Lineage Structure\label{sec: least_time_principle}}
The geometry of an advancing population front in a landscape of hotspots can \review{be approximated as a wavefront whose dynamics can be determined using} an analogy to light rays propagating through a heterogeneous medium, with hotspots effectively behaving as diverging lenses ~\cite{Moebius2021}. Building on this analogy, we demonstrate here that geometrically determined paths of least travel time through a landscape of hotspots predict much of the observed \changed{genetic structure of pinned lineages}, including which founding ancestors will have surviving descendants.

\begin{figure*}[hbt]
    \centering
    \includegraphics[width=0.6\linewidth]{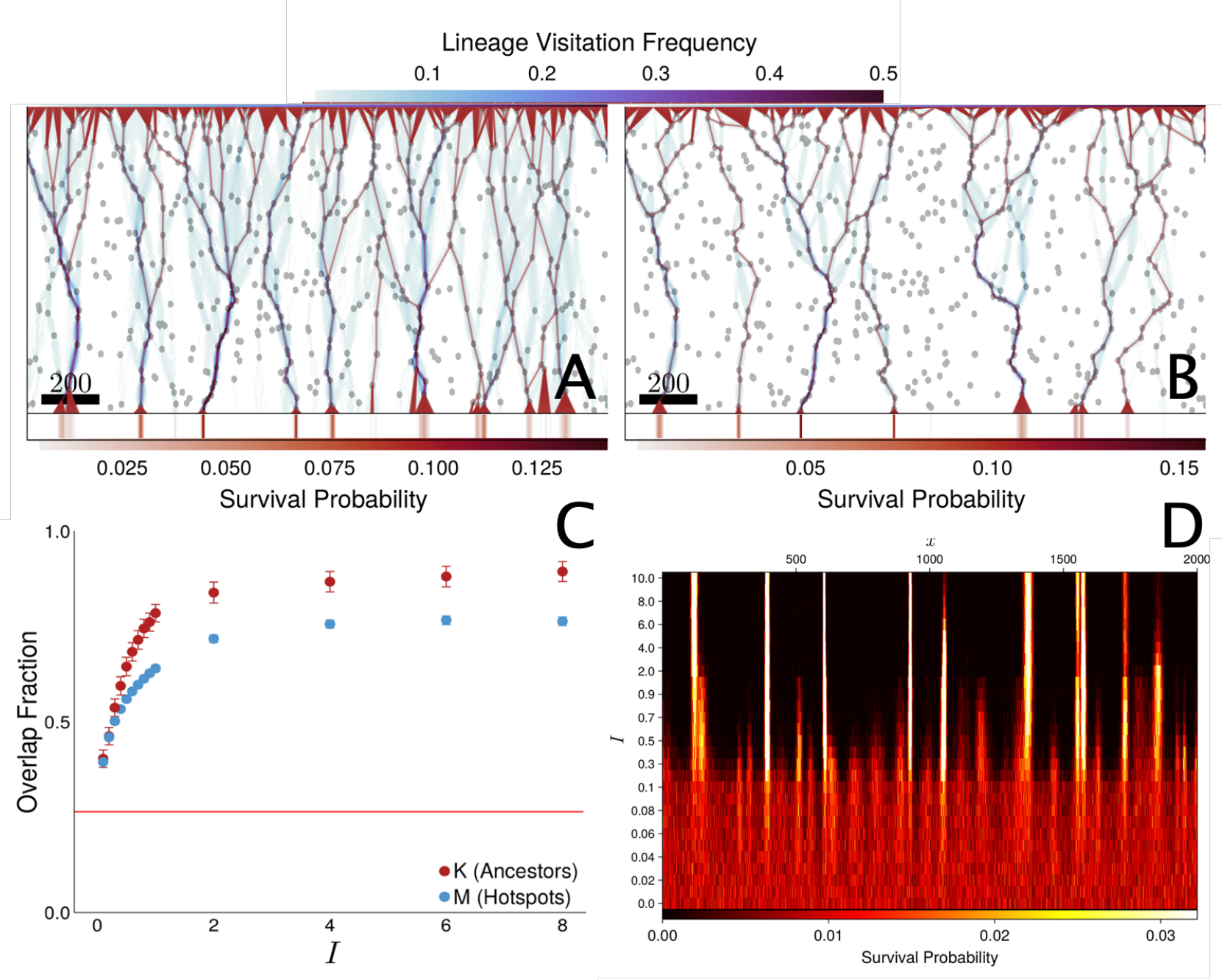}
    \caption{Environmental pinning at the end of range expansions starting from a line of occupied sites. (A, B) lineage site-visitation frequency maps at hotspot area fraction $\phi=0.1$ and hotspot size $R = 10$ with (A) hotspot intensity $I=1$ and (B) $I=10$ for the same landscape. Hotspots are indicated in grey. Ancestor survival probabilities as functions of lateral position are shown beneath.  \review{Continuous red lines indicate all paths whose net travel times are within 6\% of the fastest path for each individual at the front.}
    (C) \review{Overlap fraction between lineages and fastest paths ($M$, blue markers) as well as between fastest path terminal positions and surviving ancestors ($K$, red markers), obtained from averaging $k$ over 20 landscapes and 100 independent runs per landscape. Error bars represent standard error. Horizontal (red) line indicates the lower bound estimate for $M$ in the presence of ``no-influence'' hotspots as described in the main text.}
    (D) \review{Ancestor survival probabilities as a function of ancestor position for the hotspot landscape of (A, B) at a sequence of hotspot intensities}.
    }
    \label{fig:heatmapLineages}
\end{figure*}

\changed{
Pinned lineages sequentially visit a set of favored hotspots as they traverse the landscape. To focus on the trade-offs that generate these favored paths, we ignore the stochastic wandering of lineages in the regions between hotspots. Instead, motivated by known connections between optimal paths in random media and Eden model lineages~\cite{Roux1991,Cieplak1996,Manna1996,Cieplak1999}, we approximate each lineage as a contiguous sequence of line segments that connects an individual at the population front with the ancestors at $h=0$ by passing through the centers of hotspots visited by the lineage. Each segment adds to the travel time by an amount proportional to its length, reduced by the time saved in going through a hotspot compared to the background environment. Under this approximation, the lineage of an individual at the front is the sequence of line segments that connects the individual to one of the ancestors at the bottom edge of the range, while minimizing the total travel time. These sequences, which we term the \emph{fastest paths}, are computed using the Floyd-Warshall algorithm~\cite{Cormen2001}; see \textbf{SI Appendix} for details.
At low hotspot intensities, fastest paths run vertically down from population-front source to ancestor to minimize the Euclidean distance traversed (SI Appendix Fig~11). At high hotspot intensities, the reduction in travel time obtained by passing through hotspots induces paths to wander laterally to visit more hotspots. The fastest paths through a particular landscape result from a trade-off between minimizing the distance traversed and maximizing the number of hotspots visited.

Fig.~\ref{fig:heatmapLineages}\textit{A},\textit{B} show fastest paths (red continuous lines) overlaid with lineage site-visitation maps for hotspot intensities $I=1$ and $I=10$, respectively.
\review{Near $h=0$, any pinned lineage becomes diffuse in the region below the earliest encountered hotspot, reflecting a narrow range of ancestors that all have roughly equal likelihood of reaching that hotspot first. To reflect this degeneracy when comparing the fastest path model to the Eden lineages, 
we identify, for each individual node at the front, the set of optimal paths connecting that node to the row of ancestral nodes at $h=0$.
From this set, we collect all paths whose net travel times lie within 6\% of the smallest travel time and include them in the ensemble of fastest paths. 
As a result, each frontier individual contributes multiple fastest paths to the ensemble, but these paths almost always differ only in the last step from the lowest hotspot to the $h=0$ population (except for rare instances of nearly-degenerate optimal paths through the landscape which are unlikely for well-separated hotspots).
This results in the emergence of triangular regions at the base of the fastest paths.}
Below these site-visitation maps, we also show survival probabilities for ancestral demes as one-dimensional heatmaps indicating the probability that a given site has offspring at the front at the end of the simulation. Pinning of lineages to hotspots is visible at both $I$ values. However, lineage probability distributions are much more concentrated near fastest paths for $I=10$, which we describe as a ``strong'' pinning regime, compared to ``weak'' pinning at $I=1$. \review{Fig.~\ref{fig:heatmapLineages}\textit{D} shows the transition from weak pinning to strong pinning.}

To quantify the degree to which fastest paths predict lineage structure and surviving ancestors, we examine how commonly fastest paths coincide with lineages with regard to visited hotspots and with regard to lineage root positions. First, we assess to what extent the subset of hotspots through which lineages most often pass, as seen in Fig.~\ref{fig:landscapeView} and Fig.~\ref{fig:hotspot_pinning}, are a result of chance or arise due to fastest paths passing through them. To that end, we calculate the overlap fraction of visited hotspots $m(G,S):={|G \cap S|}/{|G \cup S|}$ where $G$ and  $S$ are the subset of hotspots visited by lineages and fastest paths in an individual simulation, respectively, and $|\cdot |$ represents the number of elements in a set. $M=\left\langle m \right\rangle$ denotes the ensemble mean,  obtained by averaging over 20 landscapes each with 100 independent simulations. Thus, $M=1$ when fastest paths and lineages pass through the same subset of hotspots. A lower bound for $M$ is the null expectation at $I=0$ for the coincidental overlap of lineages with ``hotspots'' that exert no influence. We calculate this lower bound for all intensities by randomly generating 20 additional different distributions of hotspots with identical size and area fraction and computing \review{$M$ from these no-influence ``hotspots''} (red line in Fig.~\ref{fig:heatmapLineages}\textit{C}).

Shown in Fig.~\ref{fig:heatmapLineages}\textit{C} is $M$ versus hotspot intensity $I$, with uncertainty bars calculated as standard error. As expected, for low intensities there is little overlap between the fastest paths and the full lineages, which are dominated by demographic noise at the level of individual lattice sites. However, above a threshold intensity ($I \approx 2$ for these parameters), 75\% of the hotspots visited by lineages are correctly predicted by fastest paths, even though the latter was computed without incorporating the lattice of demes that contribute to the stochastic meandering of lineages in the Eden model simulations. The effect of saturation at high intensities is explained by the fact that the travel times depend on intensities via the quantity $I/(I+1)$ when hotspots are well-separated (see \textbf{SI Appendix}).

Upon comparing lineages to fastest paths in Fig.~\ref{fig:heatmapLineages}\textit{B}, we observe that the lattice-level wandering causes lineages to make excursions to hotspots near a dominant pinned path, but lineages often return to the dominant path and can be traced back to one of the founding ancestors with high survival probability.
As a result, we expect that the agreement of founding ancestors between lineages and fastest paths will be even higher than the overlap of hotspots. 
For a given landscape and simulation instance, we define $k$ as the fraction of lineages \review{which lie within a triangular region around the base of fastest paths as described above.} 
The ensemble average of $k$, averaged over 20 landscapes each with 100 independent simulations, provides the ancestor overlap statistic, $K \equiv \langle k \rangle$.
Shown in Fig.~\ref{fig:heatmapLineages}\textit{C} is $K$ versus hotspot intensity $I$ at fixed hotspot area fraction and hotspot size, averaged across 20 landscapes. At $I=0$ fastest paths become shortest paths and hence every position $(x,0)$ is visited, resulting in $K=1$.
We observe that $K$ increases with $I$, approaching 90\% in the strong pinning regime. 
These results show that deterministic fastest paths computed using only the hotspot positions and intensities can predict most of the surviving ancestry, without needing details of short-distance stochastic effects.
}

\subsubsection*{Meandering of Lineages and of Fastest Paths\label{sec: demographics_fastest}}

\begin{figure}[tb]
    \centering
    \includegraphics[width=0.49\textwidth]{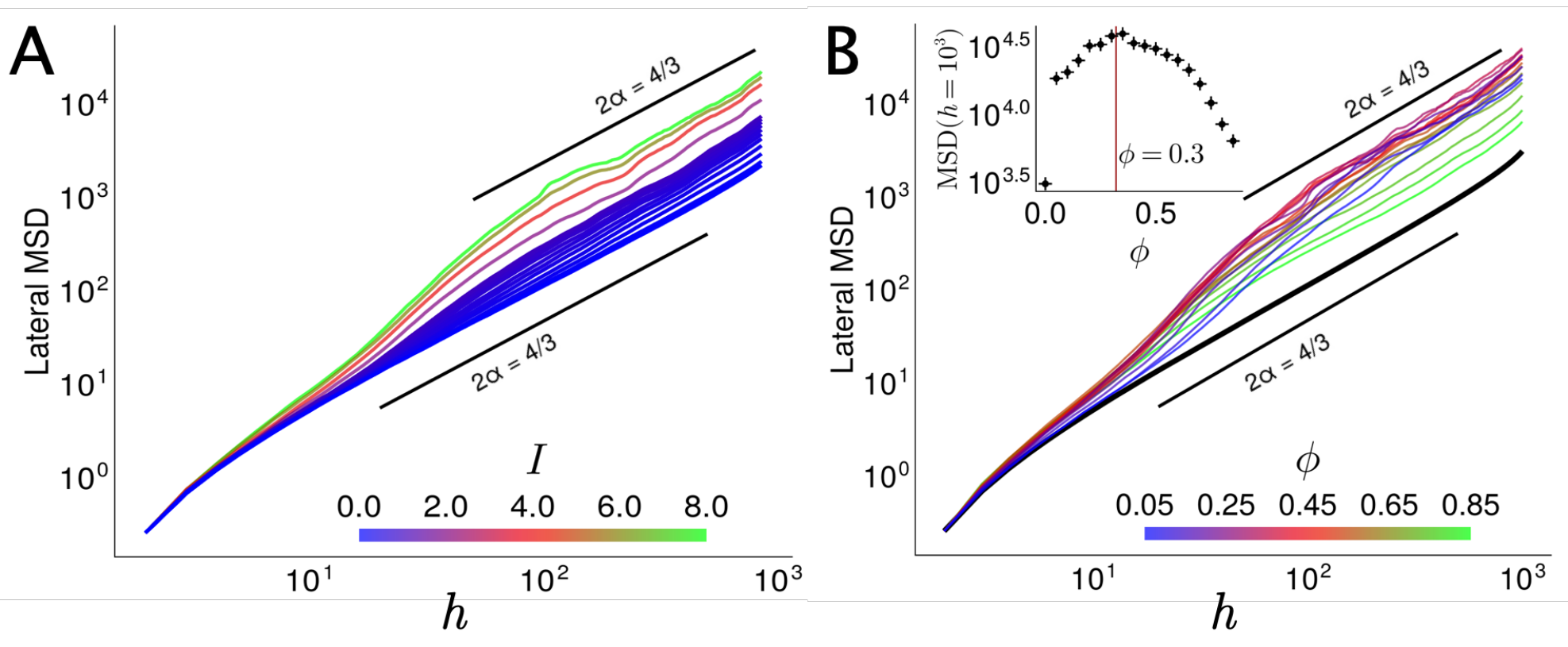}
    \caption{\review{Lateral mean-squared displacement (MSD) of lineages as a function of vertical expansion distance $h$ at (A) various $I$ and fixed $\phi=0.1$ and $R=10$, and (B) fixed $I=8$ at various $\phi$. Reference lines show power-law scaling $\mathrm{MSD} \sim h^{2\alpha}$ corresponding to KPZ superdiffusive wandering (slope $2\alpha=4/3$)}. Reference curve for uniform landscape is highlighted in black in (B). \review{Inset in (B) shows the MSD at the maximum height $h=10^3$. Vertical brown line indicates $\phi = 0.3$.}
    }
    \label{fig:MSD}
\end{figure}

\review{In \textbf{Sector coarsening}, we showed that the loss of diversity in the presence of hotspots was controlled by an exponent $\alpha\approx 2/3$, consistent with prior expectations for \emph{uniform} landscapes where the coarsening was governed by KPZ statistics. 
The statistics of sector coarsening are closely} related to the lateral meandering of lineage\review{s through the landscape}, as lineages coalesce whenever they intersect \review{and each remain inside one sector} \cite{Gralka2016, Chu_2019}.  \review{Thus, with increasing expansion distance $h$, growth of} lateral mean-squared \review{lineage} displacement \review{as $(\text{MSD}) \sim h^{2 \alpha}$ corresponds to sector size \review{growth with scaling} $l_s \sim \sqrt{\text{MSD}} \sim h^{\alpha}$.
The estimate $\alpha \approx 2/3$ from Fig.~\ref{fig:sectorSizes} therefore predicts that the lateral MSD of lineages grows superdiffusively; i.e.\ faster-than-linearly with expansion distance}. In this section, we examine how  environmental heterogeneity influences this lineage meandering in order to gain insight into the accelerated loss of genetic diversity. 

\changed{
Fig.~\ref{fig:MSD}A shows the forward-in-time lateral MSD of lineages versus expansion distance $h$ for a range of intensities at an intermediate hotspot density. 
At expansion distances that are large compared to the lattice size yet small compared to the typical hotspot separation $\lambda$, lineages have encountered few hotspots and the meandering is identical to that on a uniform landscape, which approaches the KPZ expectation $\text{MSD} \sim h^{2\alpha}$ with $\alpha = 2/3$.
As lineages encounter more hotspots at larger $h$, the transverse wandering is enhanced relative to the uniform landscape with a stronger enhancement for higher hotspot intensities (Fig.~\ref{fig:MSD}A), consistent with our observation of enhanced sector coarsening induced by hotspots.
In the strong pinning regime, $I \gg 1$, the lateral MSD growth again approaches superdiffusive growth, consistent with the KPZ wandering exponent $\alpha = 2/3$.
Interestingly, the deterministic lineage trajectories dictated by strong pinning to a structured landscape are thus governed by a  wandering exponent similar to that of demographic-noise-dominated lineages in a uniform landscape, even though the sources of the underlying stochasticity are very different. 
Our model of strongly pinned lineages as fastest paths through the hotspot landscape suggests a potential explanation for this fact: the fastest paths can be mapped to the contours of a directed polymer in a random medium (DPRM) at zero temperature, as we show in the \textbf{SI Appendix}, and related to the Brownian polymer model in a Poisson point random potential\cite{Comets2005}.
A wide range of DPRM models with different random potentials have been shown to belong to the KPZ universality class~\cite{Corwin2011}; therefore, we hypothesize that the  fastest path model described and characterized above also belongs to the same universality class.
In \textbf{SI Appendix}, we provide additional numerical evidence that the fastest paths exhibit lateral wandering characterized by an exponent $\alpha=2/3$ over a range of hotspot densities. 

In Fig.~\ref{fig:MSD}B, we report the behavior of the lateral fluctuations of lineages at high hotspot intensity over a range of hotspot area fractions.
We find that the wandering for large $h$ is consistent with $\alpha = 2/3$ over a wide range of hotspot area fractions.
The size of the fluctuations \review{at large $h$} increases with area fraction up to around $\phi = 0.3$, and then decreases \review{(Fig.~\ref{fig:MSD}B Inset)}. This non-monotonic dependence of lineage meandering on $\phi$ is understandable in light of the fact that $\phi=1$ represents a uniform landscape, identical to $\phi=0$ except for a faster front propagation speed. More broadly, when the hotspot area fraction is increased beyond the percolation threshold for overlapping disks (here, hotspots) at $\phi\approx 0.68$ \cite{Xia1988}, lineages can traverse the range without leaving hotspots, so the environment becomes a medium of increased reproduction rate punctuated by ``cold spots'' of lower reproduction rate. We note, however, that the greatest lineage wandering occurs not at $\phi \approx 0.68$ but, mysteriously, at the much smaller area fraction of $0.3$.
As $\phi$ increases above $0.3$, the lineage-pinning effect of the environment is diminished, and lineage wandering weakens as it returns to being driven by demographic noise rather than by hops between hotspots. 
}

\subsubsection*{Lineage Coalescence Rate and Common Ancestry\label{subsubsec:commonancestry}}
\begin{figure*}[h]
    \centering
    \includegraphics[width=0.99\textwidth]{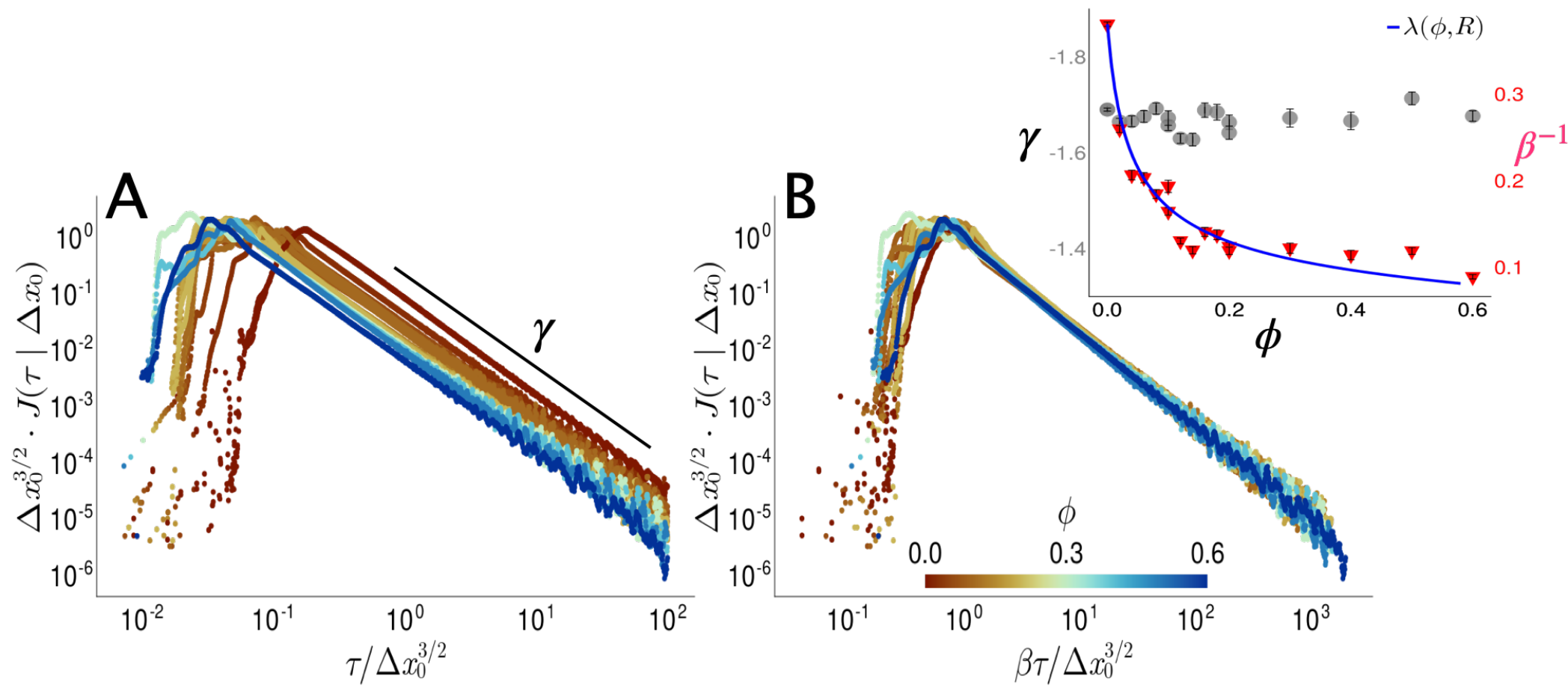}
    \caption{Log-log plot of the scaled lineage coalescence rate, $\Delta x_0 ^{3/2} J(\tau \ \vert \ \Delta x_0)$, for reverse-time $\tau$ and final-time separation $\Delta x_0$, at $I=3$, $R=10$ and for various $\phi$, plotted against (A) $\tau / \Delta x_0^{3/2}$ and (B) with $\beta \tau / \Delta x_0^{3/2}$. \review{The datasets for each $\phi$ include a range of $\Delta x_0$ values as described in the main text.}
    The horizontal colorbar indicates hotspot area fraction for the corresponding coalescence curves. Inset shows $\gamma$ (circles) and $\beta^{-1}$ (triangles) calculated from a linear fit of $\log(\Delta x_0 ^{3/2} J)$ vs.\ $\log(\beta \tau / \Delta x^{3/2})$ for $\tau / \Delta x_0 ^{3/2} \gg 1$. \review{Also shown is the typical hotspot separation length scale $\lambda(\phi, R)$ (blue curve) scaled to the maximum value of $\beta^{-1}$.}}
    \label{fig:coalescence}
\end{figure*}

One important measure of the biological consequences of environmental lineage pinning is provided by the expected time $T_2$ since common ancestry of two sampled organisms in a population. In population genetics, $T_2$ is proportional to the expected number of pairwise nucleotide site differences between two sampled genomes, assuming a constant rate of neutral mutations~\cite{Kimura1969a}. \review{With} lineages \review{viewed} as random motion through the environment, features of genetic structure are illuminated by  the coalescence rate $J(\tau \ \vert \ \Delta x_0)$ representing the probability per unit time that lineages ending at two sites separated by lateral distance $\Delta x_0$ on the population front have a most recent common ancestor born at reverse-time $\tau$.  

A mapping of lineage motion to the theory of random walks has resulted in estimations for the asymptotic power laws in range expansions in uniform environments \cite{Chu_2019}: It was found that $J$ depends on $\tau$ only through the combination $\Delta x_0 / \tau^\xi$ with KPZ exponent $\xi = 2/3$. (Here and throughout, lengths are implicitly scaled by deme size and by the inverse of the intrinsic reproduction rate $r_0^{-1}$.) In particular, in the regime $\tau / \Delta x_0 ^{3/2} \gg 1$, which represents coalescence events in which lineages located nearby at the front can be traced back to a distant common ancestor, has a form 
\begin{equation}
    J(\tau \ \vert \ \Delta x_0) \propto \frac{1}{\Delta x_0 ^{3/2}} \left( \frac{\tau}{\Delta x_0 ^{3/2}} \right)^\gamma,
\end{equation}
with a numerically determined exponent $\gamma = -1.64 \pm 0.05$ for uniform landscapes \cite{Chu_2019}.

Following Ref.~\cite{Chu_2019}, we calculate $J$ for all lineage pairs with separation $\Delta x_{0}$ at the front less than $\Delta x_{0,\mathrm{max}}$, which is kept much smaller than the system width $L$ to avoid finite size effects. \review{(}As expected, we find that when $\Delta x_{0,\mathrm{max}}$ is less than the hotspot diameter $2R$, the coalescence rate $J(\tau \ \vert \ \Delta x_0$) reproduces the results in Ref.~\cite{Chu_2019}.
For this reason, we set $\Delta x_{0,\mathrm{max}} = 300$ hereafter so that 
$2R < \Delta x_{0,\mathrm{max}}  \ll L$; other $\Delta x_{0,\mathrm{max}}$ values in this regime produced similar results.\review{)} With \review{the choice $\Delta x_{0,\mathrm{max}} = 300$}, we find that the coalescence rate \review{$J(\tau | \Delta x_0)$ } takes on additional variation\review{, compared to the smooth variation at $\phi=0$,} in the regime $\tau / \Delta x_0 ^{3/2} \ll 1$ (Fig.~\ref{fig:coalescence}\textit{A}). Even so, the $\tau / \Delta x_0 ^{3/2} \gg 1$ regime retains its power-law form with similar exponent $\gamma$  measured for various hotspot area fractions, albeit with a shift in the distribution. 

\review{As we have found that the strongly pinned lineages wander, at large length scales, similarly to superdiffusive lineages in uniform landscapes, their}  linear coalescence dynamics ought to resemble the uniform landscape case with some time-rescaling $\beta$,
\begin{equation}
     J(\tau \ \vert \ \Delta x_0) \propto \frac{1}{\Delta x_0 ^{3/2}} \left( \frac{\beta \tau}{\Delta x_0 ^{3/2}} \right)^\gamma. \label{eq:Jbeta}
\end{equation}
We estimate $\beta$ and $\gamma$ as a function of $\phi$ via a linear fit with slope $\gamma$ in the $\tau / \Delta x_0 ^{3/2} \gg 1$ regime of the log-log plot. The constant of proportionality, $c$, omitted from Eq.~\ref{eq:Jbeta} is assumed to be independent of $\phi$, and appears in the intercept of the fit as $\log(c\beta ^{\gamma})$, with $c$ then calculated in \review{the} absence of hotspots.
The inset of Fig.~\ref{fig:coalescence}\textit{B} shows the $\gamma$ (gray circles) and $\beta^{-1}$ (red triangles) obtained in this way. \review{We find that $\gamma$ is approximately constant as $\phi$ varies, consistent with the value found in Ref.~\cite{Chu_2019}} without hotspots. 
\review{Moreover, re}plotting  \review{$J(\tau | \Delta x_0)$} with time rescaled by factor $\beta$ collapses the \review{data} onto the uniform-environment distribution \review{(Fig.~\ref{fig:coalescence}\textit{B})}. 
\review{Interestingly, we find} the time rescaling $\beta$ to be inversely proportional to the typical hotspot separation length scale $\lambda$ given by Eq.~\ref{eq:lambda} (blue curve \review{in Fig.~\ref{fig:coalescence} Inset}) since after normalization $\beta^{-1}(\phi)$ obtained through fitting are well described by $\lambda(\phi,R)$.

For our finite-time simulation, given that two sampled organisms at the population front at time $t_{\mathrm{max}}$ have a common ancestor in the expansion history, the expected time $T_2$ since their common ancestry is calculated from the first moment of $J$ as
 \begin{equation}
    T_2(\Delta x_0 ) = \frac{\int_0^{t_{\mathrm{max}}} \tau \cdot J(\tau \ \vert \ \Delta x_0) d\tau}{\int_0^{t_{\mathrm{max}}} J(\tau \ \vert \ \Delta x_0) d\tau}. \label{t2def}
\end{equation}

\begin{figure}[bt]
    \centering
    \includegraphics[width=0.47\textwidth]{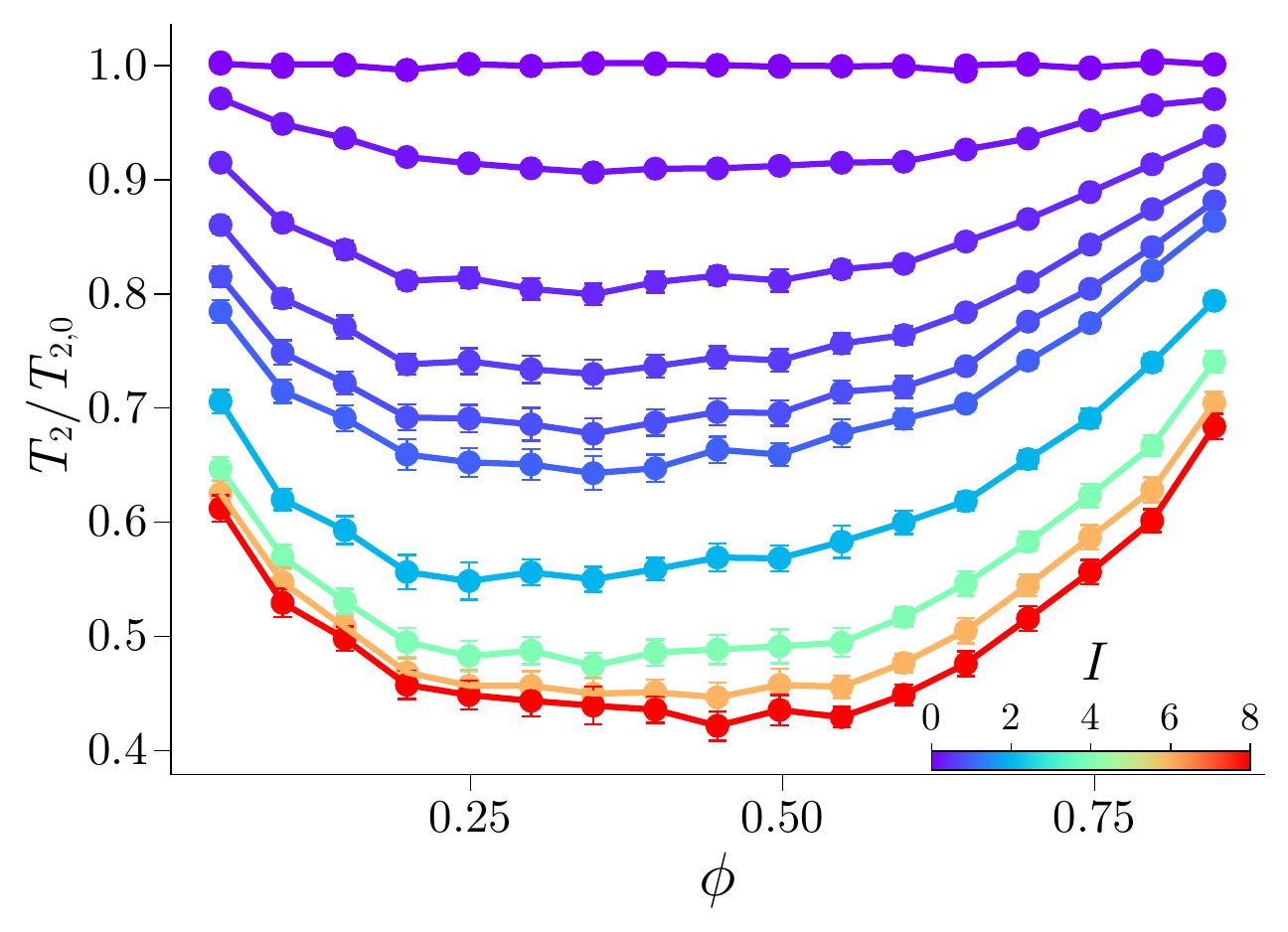}
    \caption{Mean common ancestry time $T_2$, normalized by that of the uniform landscape $T_{2,0}$, vs.\ hotspot area fraction $\phi$ at different hotspot intensities $I \in [0, 0.1, 0.2, 0.3, 0.4, 0.5, 0.6, 0.7,0.8,0.9, 1, 2, 4, 6, 8]$.}
    \label{fig:T2VSDensity}
\end{figure}

We calculate $T_2$ from Eq.~\ref{t2def}, averaged over the entire final-time population front and normalized to \review{the same} measure calculated in the uniform landscape, $T_{2,0}$. 
Fig.~\ref{fig:T2VSDensity} shows $T_2/T_{2,0}$ plotted against $\phi$ for several hotspot intensities $I$. Notably, $T_2/T_{2,0}$ exhibits a non-monotonic dependence on hotspot area fraction $\phi$. At low $\phi$ below $\phi \approx 0.3$,  adding more hotspots shifts lineage coalescences toward more recent common ancestors, on average\review{,} and $T_2$ declines.
However, for $\phi$ values greater than $\phi \approx 0.55$, $T_2/T_{2,0}$ increases roughly linearly with $\phi$. \review{The trend of decreasing $T_2/T_{2,0}$ with increasing hotspot intensity $I$ appears to saturate:   reductions in normalized mean common ancestry times are similar for $I=6$ and $I=8$ over all $\phi$ values.} 

\review{The measure $T_2/T_{2,0}$ is equal to 1 not only at $\phi=0$ (by definition) but also at $\phi=1$ because the landscape is then}
characterized by a uniformly increased replication rate and front propagation speed. 
\review{A possible explanation for the transition in $T_2/T_{2,0}$ between the low $\phi$ and high $\phi$ behavior is the duality in which, at high $\phi$, non-hotspot areas stop percolating and begin to behave as isolated growth obstacles (regions of decreased replication rate) which lineages now \review{tend} to avoid. This return to statistical behavior of uniform landscapes is also evidenced by the lateral fluctuations of lineages, which reach a maximal value and then decrease with increasing $\phi$ above a critical value $\phi \approx 0.3$ (Fig.~\ref{fig:MSD} inset). }

\section*{Conclusion}\label{sec:discussion}
We have found that an environmental \emph{pinning} effect on the lineages arises in structured landscapes of hotspots, which we can understand as superdiffusive random walks that are biased toward certain optimal paths. Using the \review{Floyd-Warshall} algorithm and viewing hotspots as geometric lenses of relative index of refraction $v_{0} / v_{h} = (I + 1)^{-1}$, we found that such optimal paths correspond to fastest paths that light would take in the analogous system. For large hotspot intensity $I$, these fastest paths predict much of the lineage motion, including the survival probability of ancestral lineages. In such systems with strong environmental or landscape-induced noise, measures of sector coarsening reflect the intrinsic, demographic noise at short timescales and cross over to a \emph{different} KPZ-like regime determined by the environment at large timescales. A non-monotonic behavior was observed in the mean time $T_2$ since common ancestry for pairs of demes at the front as hotspot area fraction $\phi$ was varied, indicating that $0.3 \lesssim \phi \lesssim 0.6$ produces fronts with the least genetic diversity. 

A key manifestation of the competition between demographic and environmental noise is highlighted by our comparison of simulated lineage positions and calculated fastest paths. Viewed geometrically, there is an optimal path trade-off that arises from the tendency of demographic noise to favor fluctuations around shortest-distance paths (Fig.~\ref{fig:landscapeView}\textit{A}) while environmental noise pulls lineages into excursions through a consistent subset of hotspots that increase the travel distance but decrease the travel time  (Fig.~\ref{fig:heatmapLineages}\textit{A,B}). At low hotspot intensity, demographic noise dominates and pulls lineages into shortest paths, resulting in many lineage tracks located away from fastest paths, as well as low overlap between hotspots visited by lineages and fastest paths in Fig.~\ref{fig:heatmapLineages}\textit{A}. At the other extreme, high-intensity hotspots cause lineages to follow fastest paths with high probability (Fig.~\ref{fig:heatmapLineages}\textit{B}). In this strong pinning regime, lineage meandering is dictated almost entirely by that of the fastest paths (Fig.~\ref{fig:heatmapLineages}\textit{B})
At \review{sufficiently large expansion distances}, the lineages exhibit the \review{superdiffusive} wandering exponent of the KPZ class, similar to lineages in uniform environments. 

What does this mean for the genetic structure? Individual simulation snapshots may give the impression that growth through a landscape of hotspots produces similarly random lineage trees to those in uniform environments. However, we have found that a structured environment of high-intensity hotspots pins lineages to fastest paths, giving a strong deterministic component to the genetic structure of the range expansion for a given environment. This lineage pinning is associated with an accelerated loss of genetic diversity. Furthermore, sites (or more generally individuals) found near one of the environmentally-determined optimal paths have the highest likelihood of influencing future genetic composition through surviving descendants.  Our results suggest that, for non-neutral evolution, random hotspot landscapes may increase the importance of chance relative to selection\review{,} as was seen \review{for individual obstacles \cite{Moebius2015} and} in landscapes of obstacles \cite{Gralka2019}. Future studies could examine whether beneficial mutations are more frequently lost from the expansion when they happen to originate away from a fastest path.

In summary, KPZ statistics seen in uniform landscapes can also arise in structured environments, coming from demographic noise or from environmental quenched random noise. In the absence of environmental heterogeneity, demographic noise alone determines the outcome of neutral evolution. Low\review{-}intensity hotspots give rise to fastest paths through the landscape but are too weak to out-compete lineage superdiffusive motion \review{caused by demographic noise}. More intense hotspots impart a sufficiently high boost in reproduction rate to portions of the population front entering a hotspot that they consistently out-compete the population remaining outside of a hotspot, producing lineages bound close to the subset of hotspots defining a landscape's fastest paths. In this lineage-pinning regime, the genetic structure of the range expansion becomes predictable, to a large degree, from the geometry of the hotspot distribution. In other words, the quenched noise of the environment replaces demographic noise as the main determinant of evolutionary dynamics.  

\subsection*{Acknowledgements}

J.P.\ thanks Abhijeet Melkani for discussions about the diffusive wandering of directed polymers. J.G.N. acknowledges support from the National Science Foundation Research Traineeship in Intelligent Adaptive Systems (NSF-DGE-1633722) and the NSF-CREST Center for Cellular and Biomolecular Machines at UC Merced (NSF-HRD-1547848). J.G.N.\ and D.A.B.\ acknowledge support from the Hellman Fellows Fund. W.M. acknowledges support by BBSRC via BBSRC-NSF/BIO grant BB/V011464/1.


\onecolumn
\section*{Materials and Methods}
    \subsection*{Gillespie Algorithm}
        We introduce environmental heterogeneity into the Eden model through a space-dependent growth rate, Eq.~\ref{eq:growth_rates}. Reproduction occurs through an implementation of the Gillespie algorithm: for a population front consisting of $n_h$ demes in hotspots and $n_0$ demes outside of hotspots, the probability of selecting a particular deme inside of a hotspot as the next site to reproduce is $P_h = r_h / Q$ with $Q = n_h r_h + n_0 r_0$ being the sum of reproduction rates of demes at the front. After replication, time is updated according to the Gillespie algorithm by $\delta t = -\log(\eta)/Q$, with $\eta$ being a uniform pseudo-random number in the range $[0,1)$ \cite{Gillespie1976}.

    \subsection*{Determination of Optimal Paths}
    Fastest paths are computed using the Floyd-Warshall algorithm implemented in the Python SciPy package~\cite{scipy}.
    
\section*{Supporting Information Appendix (SI)}

\subsection*{Parabolic Description of Sector Boundary Induced by Single Hotspot}\label{app:parabola}

\begin{figure}[hbt]
    \centering
    \includegraphics[width=0.6\textwidth]{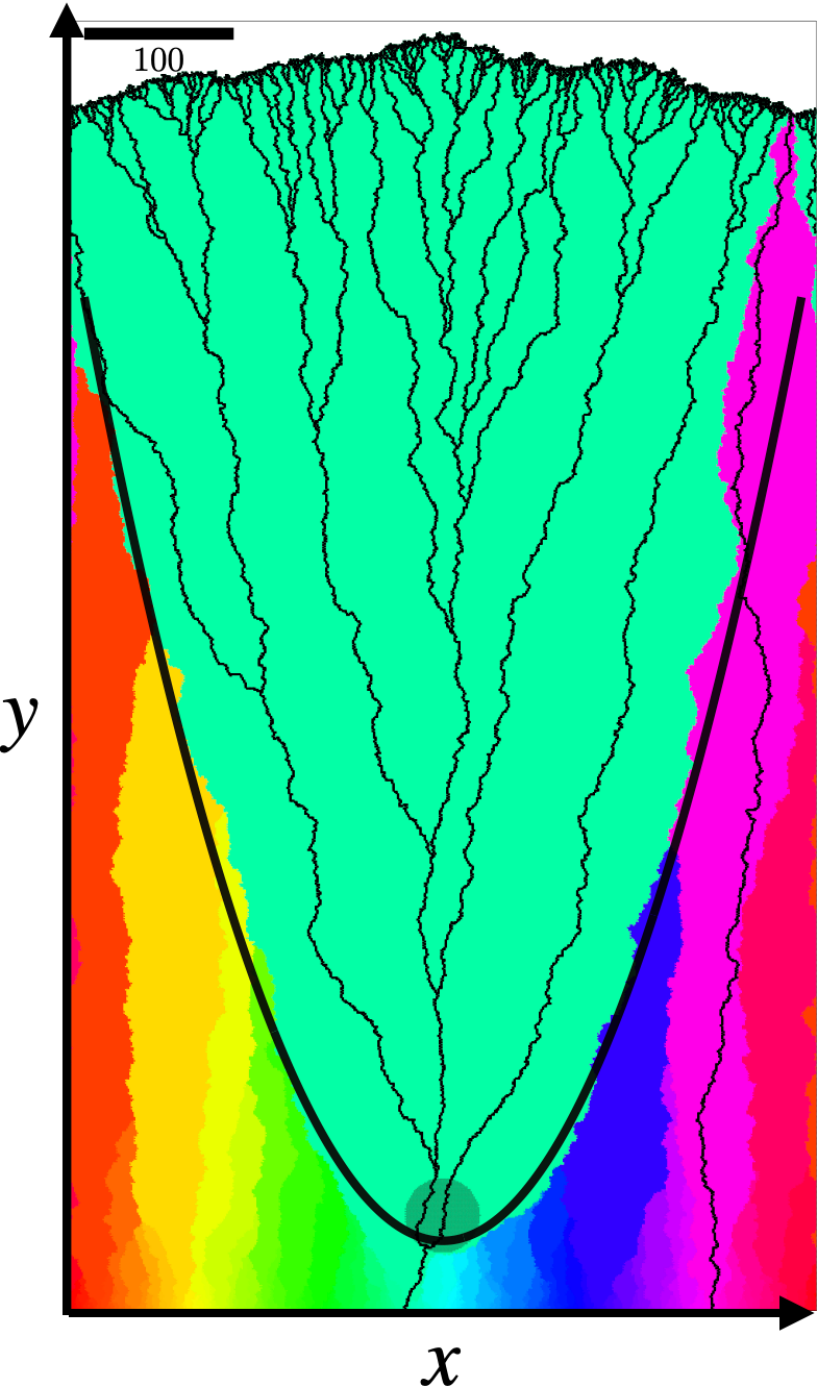}
    \caption{Illustrative example of the Eden model for a range expansion beginning as a single line at $y=0$ and proceeding through a heterogeneous landscape with a single hotspot (gray disk), and with each color representing a distinct ancestral deme. Lineages are shown as black lines. Sector boundaries formed through the influence of a single hotspot are well-described by a parabola (bold black curve) of the form in Eq.~\ref{eq:parabola_form}. Here the hotspot center is located at $(500, 50)$ with $R=25$ and hotspot intensity $I=10$.}
    \label{fig:parabola_longtime}
\end{figure}

An unperturbed plane wave and a circular wave originating at the center of the hotspot together heuristically well-describe the front shape after it passes a hotspot; see Ref.~\cite{Moebius2021} for details. The intersection points of the circular and plane waves trace out a parabola given, in the notation of this work, by 
\begin{equation}\label{eq:parabola_form}
    y = \frac{I+1}{4IR} (x - h_x)^2 + h_y - R\frac{I}{I+1},
\end{equation}
where $R$ is hotspot radius, $I$ is hotspot intensity as defined in the main text, and $(h_x, h_y)$ is the position of the hotspot's center.
In the geometric optics description of front propagation, the section of the front that is a result of the circular wave is associated with fastest paths through the hotspot, while sections of the front away from the hotspot are associated with fastest paths that do not overlap with the hotspot
(Fig.~2b in Ref.~\cite{Moebius2021}). This means that the geometrical optics description holds at distances larger than the hotspot size, as shown in Fig.~1B in the main text and Fig.~\ref{fig:parabola_longtime}.

\subsection*{Fastest paths in a hotspot landscape without noise}
Here we describe the computation of fastest paths as approximations to lineages in the strongly pinned regime, and we analyze the statistics of these fastest paths.
The fastest path approximation to lineages is motivated by known connections between lineages in the Eden model and optimal paths in random media~\cite{Roux1991,Cieplak1996,Manna1996,Cieplak1999}. 
Specifically, Eden growth can be recast as the following optimization problem on the network of nearest-neighbor links connecting lattice sites:
First, each edge is assigned a random waiting time, drawn from a Poisson distribution whose mean is the expected reproduction time for a deme. 
The sequence of reproduction events that leads to the occupation of a particular site on the lattice is then given by the path on the network that minimizes the total waiting time from that site to the ancestral population.
The Eden cluster at any given time $t$ is then the set of sites whose optimal paths to the ancestral population at $t=0$ ($y=0$) have a net waiting time less than or equal to $t$, and the lineages are the optimal paths themselves~\cite{Roux1991}.

\subsubsection*{Network model of fastest paths}
As described in the main text, we construct fastest paths as approximations to the true lineages that ignore the stochastic wandering of lineages between hotspots but capture the speed-up in net reproduction time obtained by passing through hotspots. 
To do so, we set up an optimization problem on a spatial network whose nodes include the ancestral demes, the centers of the hotspots, and the front population whose genealogies we are interested in tracing back to the ancestors. 
In order to allow all possible paths from an individual at the front to an ancestral deme, we build a complete graph, i.e.\ each node has an edge to all other nodes. 
Each edge $(i,j)$  is assigned a waiting time $\tau_{ij}$ which is the net travel time associated with a line segment connecting the centers of its participating nodes. Under the ray optics assumption, $\tau_{ij}$ is computed from the length $l_{ij}$ of the link as:
\begin{equation} \label{eq:tauij}
    \tau_{ij} = \begin{cases}
        \dfrac{l_{ij}}{v_0} - 2 \dfrac{R}{v_0} + 2\dfrac{R}{v_h} = \dfrac{1}{v_0}\left[l_{ij} - 2R\left(1-\dfrac{v_0}{v_h}\right)\right], & l_{ij} > 2R \\ \\ 
        \dfrac{l_{ij}}{v_h} = \dfrac{1}{v_0}\left(l_{ij} \dfrac{v_0}{v_h}\right), & l_{ij} < 2R.
    \end{cases}
\end{equation}
Upon rescaling distances with the hotspot radius $R$ and waiting times with the time scale $R/v_0$, we obtain the non-dimensionalized version of \eqref{eq:tauij} as
\begin{equation} \label{eq:tauij_rescaled}
    \tilde\tau_{ij} = \begin{cases}
        \tilde{l}_{ij} - \dfrac{2I}{I+1}, & \tilde{l}_{ij} > 2 \\
        \dfrac{\tilde{l}_{ij}}{I+1}, & \tilde{l}_{ij} < 2
    \end{cases}
\end{equation}
where $\tilde{l}_{ij} = l_{ij}/R$ and $I = v_h/v_0-1$. \eqref{eq:tauij_rescaled} shows that the waiting times are determined by the dimensionless edge distances and a single dimensionless parameter, the hotspot intensity $I$.
The fastest path associated with a frontier node is the path connecting that node with any one of the ancestral nodes that minimizes the total waiting time.

\subsubsection*{Statistical properties of fastest paths}
\begin{figure}[hbt]
    \centering
    \includegraphics[width=0.5\textwidth]{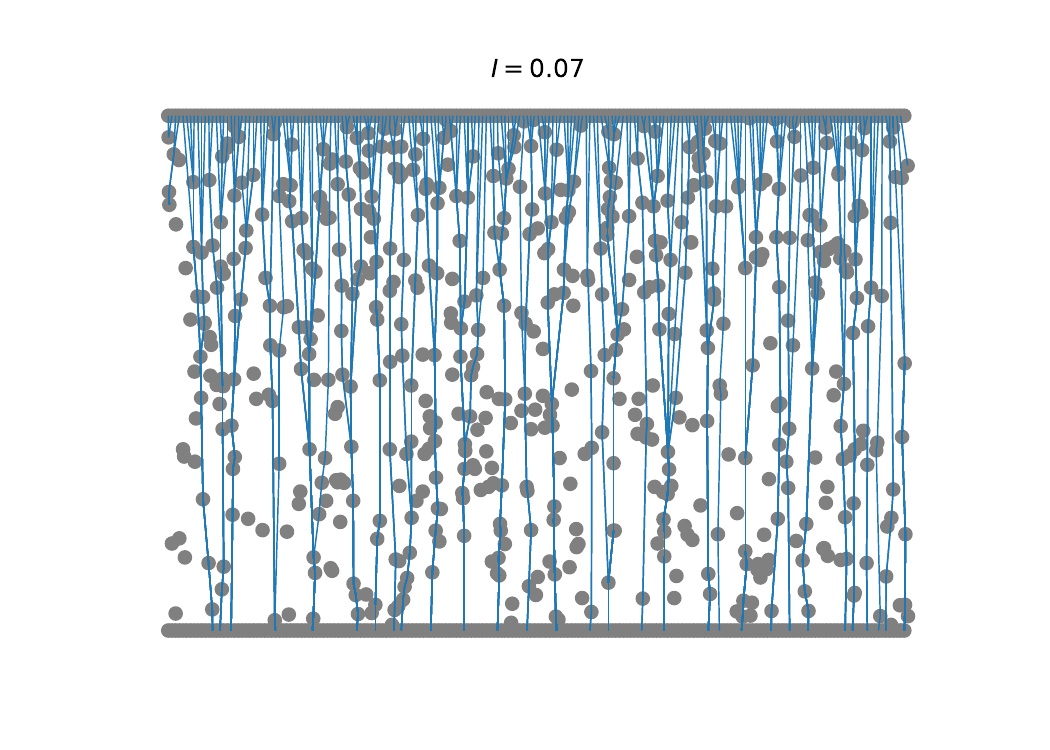}
    \includegraphics[width=0.5\textwidth]{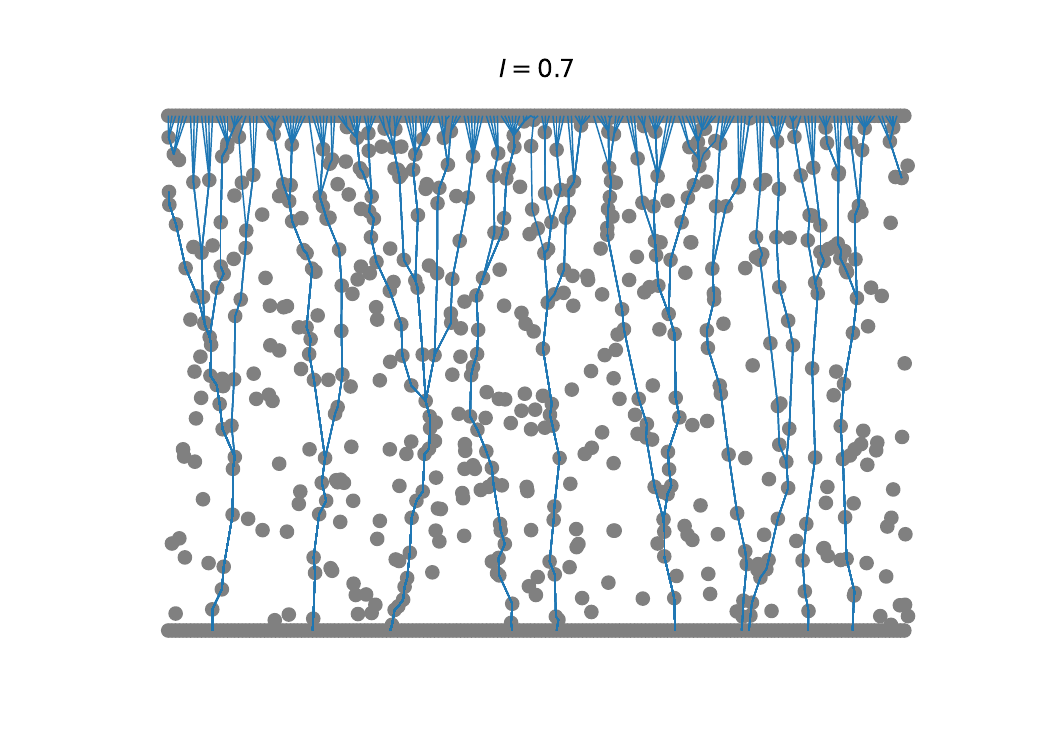}
    \includegraphics[width=0.5\textwidth]{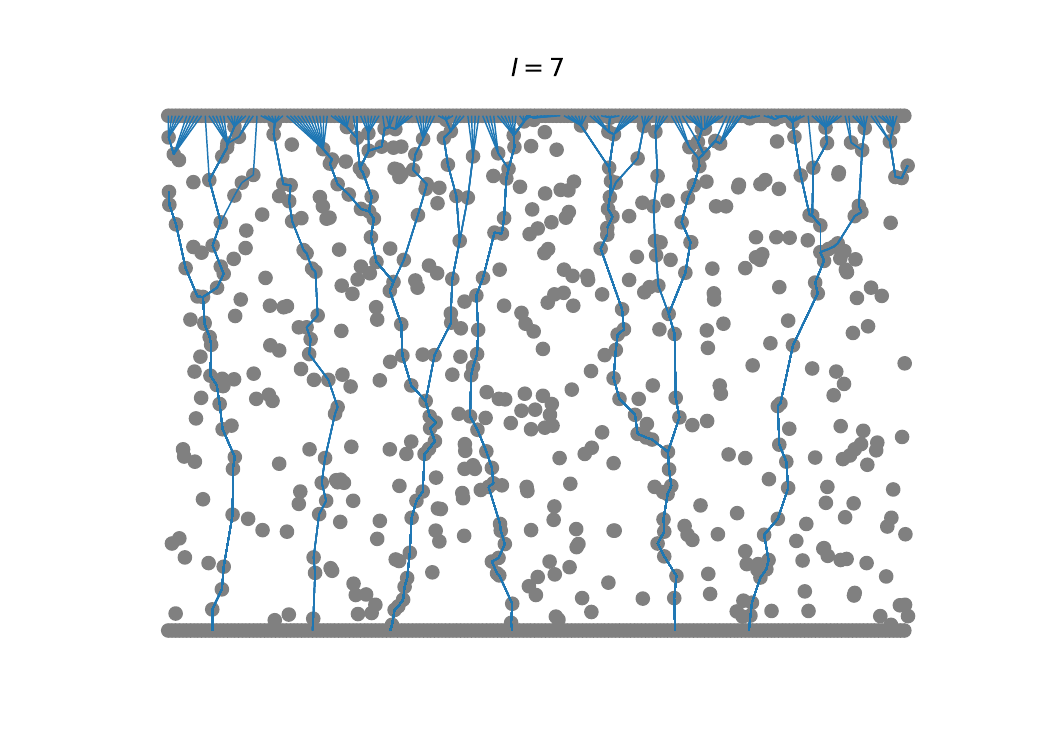}
    \caption{Fastest paths through a landscape of 587 hotspots randomly distributed in an area of extent $200R \times 100R$ (corresponding to area fraction $\phi=0.1$) for different values of hotspot intensity indicated.}
    \label{fig:fastestpaths}
\end{figure}

In Fig.~\ref{fig:fastestpaths}, we show the outcome of the fastest path computation on a fixed landscape for different hotspot intensities. 
Fastest paths are computed using the Floyd-Warshall algorithm implemented in the Python SciPy package~\cite{scipy}. 
In the limit of zero intensity, $I=0$, the fastest path associated with any frontier individual is the vertical link to the ancestor directly below it (same $x$ coordinate).
At low intensities, fastest paths remain close to this vertical line: the Euclidean distance covered by the link is the main contribution to the waiting times, and is minimized by connecting nodes with ancestors that are vertically beneath them. 
As the pinning strength is increased, paths increase their lateral wandering, picking up additional travel time so that they can accumulate the length-independent reduction in waiting time garnered by passing through hotspots.
As the fastest paths progress backwards in time, neighboring fastest paths tend to merge but not split, since the fastest-path contribution from any visited hotspot to the ancestors is unique.
Therefore, fastest paths converge to a subset of highly traversed hotspots as the originating population is approached, mirroring the lineage pinning seen in the Eden model. 

To quantify the meandering of the fastest paths in the network model, we computed ensemble averages of their lateral mean square displacements (MSDs) as a function of range height $h$, similar to the calculations for Eden model lineages in Fig.~7 of the main text. 
Specifically, we set the horizontal size of the range to $L_x= 200R$ and  generated multiple independent hotspot landscapes for each value of $h$, with area fraction $\phi$ and hotspot intensity $I$ fixed. 
The front and ancestral populations were encoded as rows of nodes horizontally spaced by $R$ at $y=0$ and $y=h$ respectively.
For each landscape, fastest paths were calculated from all population front nodes.
Since there is significant overlap among fastest paths which share a common ancestral node, we sampled one fastest path from each unique ancestral node at random and calculated the lateral mean square displacement of the path from the ancestral position.
The MSD of such paths is computed as the variance of the $x$ positions of nodes through which a fastest path passes.
By selecting all such subsets of fastest paths from 800 independent landscapes, we obtained thousands of fastest paths at each $(\phi,I,L_x,h)$ and computed the ensemble average of the MSD.

Fig.~\ref{fig:fastestpathmsd}A shows the resulting ensemble-averaged MSD as a function of expansion distance for three different area fractions with $I=7$ (strong pinning). 
The meandering is seen to be faster than the expectation for diffusing paths (which would follow $\text{MSD} \propto h$, indicated as solid line): the wandering of the fastest paths through the hotspot landscape is superdiffusive.
At large expansion distances, the MSD growth approaches the expectation from the KPZ universality class of $\text{MSD} \propto h^{4/3}$, which suggests that the statistics of the fastest paths through the landscape might also fall into the KPZ universality class.

\subsubsection*{Mapping to directed polymers}
Additional evidence supporting the possibility that fastest paths fall in the KPZ universality class comes from mapping the fastest paths to the conformations of a directed polymer in a random medium.
In the limit that hotspots are well-separated relative to their diameter ($\tilde{l}_{ij} \gg 2$ for all $(i,j)$), and assuming that paths do not double back as they advance from origin to frontier, the fastest path of a frontier node at $x_f$ in our non-dimensionalized model corresponds to single-valued functions $x(y)$ with $x(h)=x_f$ that minimize the net waiting time
\begin{equation} \label{eq:waittimeaction}
    T[x(y)] = \int_0^h \! dy\, \sqrt{1+\left(\frac{dx}{dy}\right)^2} - \frac{2I}{I+1} \times \# \left\{ \text{hotspot positions } (y_s,x_s): x(y_s) = x_s \right\}.
\end{equation}
The second term in the above equation counts the number of hotspots encountered by the path $x(y)$.
\eqref{eq:waittimeaction} can be interpreted as an action for directed polymers in a random potential defined by the Poisson-distributed hotspot locations, whose strength is given by the intensity factor $2I/(I+1)$.
The fastest paths then correspond to the zero-temperature limit of a thermally fluctuating directed polymer in the hotspot landscape, with $x$ position confined to the frontier individual at $y=h$ but allowed to terminate at any originating point along the entire horizontal axis at $y=0$ (so-called ``point-to-line'' initial conditions~\cite{Corwin2011}). 

\begin{figure}[hbt]    
\centering
    \includegraphics[width=0.7\textwidth]{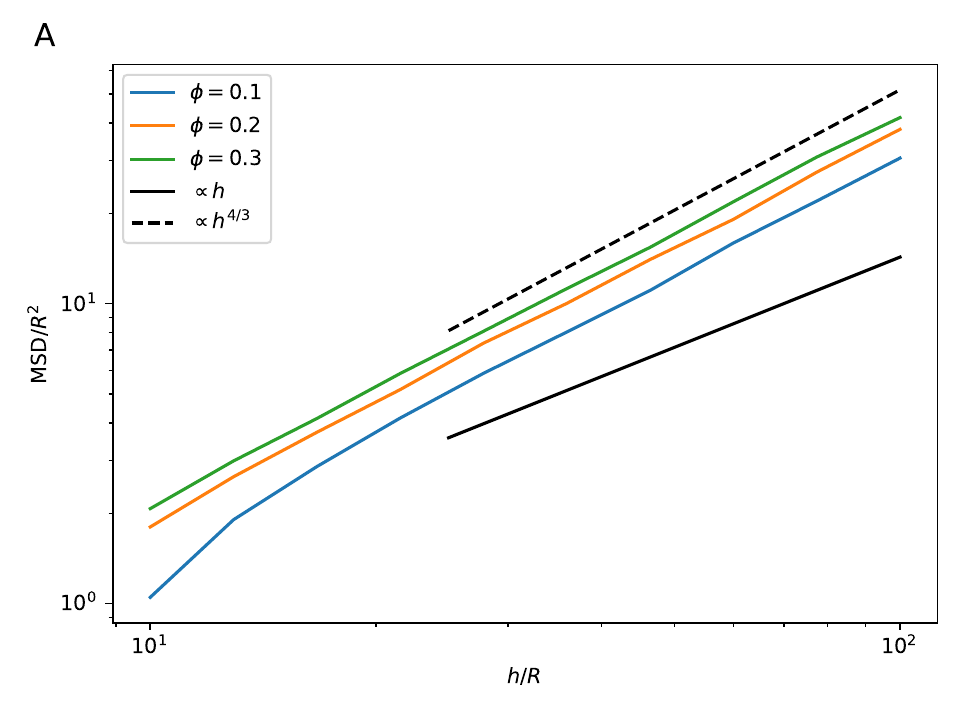} \vspace{0.2in}
    \\
\includegraphics[width=0.7\textwidth]{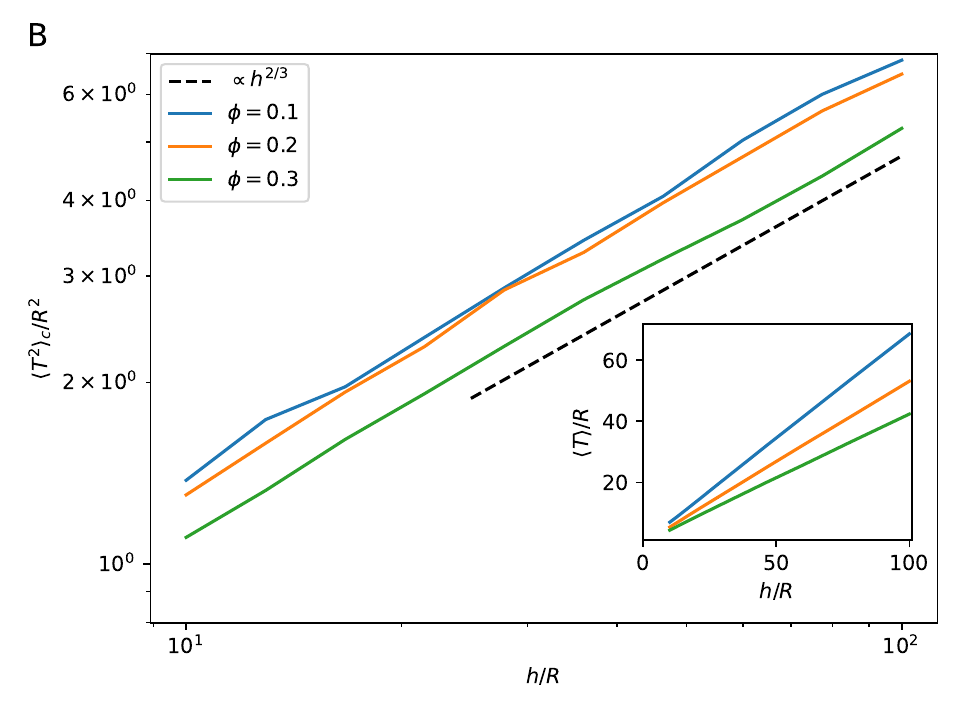}
    \caption{(A) Ensemble-averaged mean-square lateral displacement (MSD) of fastest paths through landscapes with horizontal size $200R$ as a  function of expansion distance (vertical size) $h$ at different area fractions $\phi$, with hotspot intensity $I=7$ across all parameters. Dashed line show the superdiffusive scaling predicted by the KPZ wandering exponent, $\text{MSD} \propto h^{4/3}$. For comparison, the solid line shows diffusive scaling $ \sim h$. (B) Variance in net waiting times of fastest paths as a function of vertical distance, with same landscapes and parameters as in (A). Dashed line shows the KPZ expectation $\langle T^2 \rangle_c \sim h^{2/3}$. Inset shows the linear scaling of the mean waiting time with $h$ (linear axes).}
    \label{fig:fastestpathmsd}
\end{figure}

Many directed polymer models have been shown to exhibit KPZ statistics through a mixture of numerical, approximate, and exact results; see Refs.~\cite{Halpin_Healy_1995,Corwin2011} and references therein.
A directed polymer model---the Brownian polymer in a space-time Poisson point random potential---with an action closely related to \eqref{eq:waittimeaction} was studied in Ref.~\cite{Comets2005}. 
The Brownian polymer model differs from the current model in that it constrains the path segments between the pinning points to execute Brownian walks even at zero temperature, but the effect of the Poisson random potential on net waiting times is very similar to our model. 
Ref.~\cite{Comets2005} established that the wandering of paths in the Brownian model with a Poisson random potential is superdiffusive, $\text{MSD} \propto h^{2\alpha}$ with an exponent $1 < {2\alpha} < 3/2$, and conjectured that the model belongs to the KPZ universality class, which would imply $2\alpha = 4/3$.
Our numerical results in Fig.~\ref{fig:fastestpathmsd}A show that the Poisson point random potential with straight lines between hotspots, corresponding to the zero temperature limit, generates superdiffusive paths consistent with the expectation for wandering from the KPZ universality class.
Additionally, the ensemble averages of waiting times of numerically generated fastest paths (equivalent to the free energy of the polymer at zero temperature) are shown in Fig.~\ref{fig:fastestpathmsd}B.
We observe scaling of the mean and variance of the waiting times as $\langle T \rangle \sim h$, $\langle T^2 \rangle_\text{c} \sim h^{2/3}$ which is also consistent with the behavior of the free energy of directed polymers in the KPZ universality class~\cite{Comets2005}. 

\begin{figure}[hbt]
    \centering
    \includegraphics[width=0.8\textwidth, trim={0 1mm 0 2mm},clip]{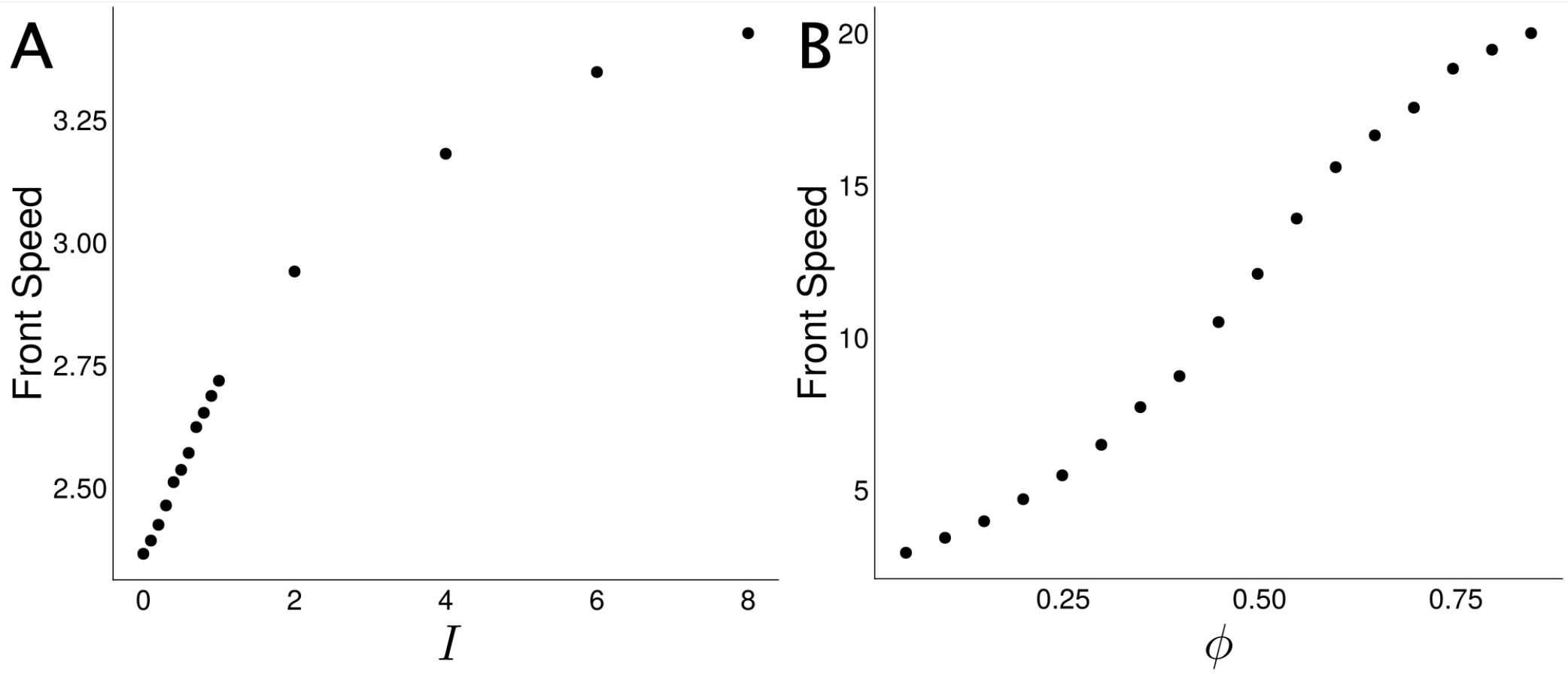} \\
    \vspace{0.4in}
    \caption{Front speeds determined from linear fits of the mean front height $h(t)$ vs.\ time $(t)$, for (A) varying hotspot intensity $I$ and fixed hotspot area fraction $\phi=0.1$ and (B) varying hotspot area fraction $\phi$ at fixed hotspot intensity $I=8$.}
    \label{fig:front speeds}
\end{figure}

\end{document}